\documentclass[aps, 10pt, twocolumn, showpacs,preprintnumbers,amsmath,amssymb,APSl,prd,nofootinbib,superscriptaddress]{revtex4-2}

\usepackage{graphicx,color}
\usepackage{subfig}
\usepackage{amssymb}
\usepackage{hyperref}
\usepackage[utf8]{inputenc}
\usepackage[english]{babel}
\usepackage{epsfig}
\usepackage{wasysym}
\usepackage{color,xcolor}
\usepackage{amsmath}
\usepackage{bm}
\usepackage{epsfig}
\usepackage{amsfonts}
\usepackage{dcolumn}
\usepackage{float}
\hypersetup{colorlinks=true, linkcolor=blue, citecolor=green}

\begin{document}

%%%%%%%%%%%%%%%%%%%%%%%%%%%%%%%%%%%%%%%%%%%%%%%%%%%%%%%%
\title{Constraints on the $\gamma$-parameter for the vacuum solution of Cotton gravity with geodesics and shadows\\}
%%%%%%%%%%%%%%%%%%%%%%%%%%%%%%%%%%%%%%%%%%%%%%%%%%%%%%%%

	\author{Ednaldo L. B. Junior} \email{ednaldobarrosjr@gmail.com}
\affiliation{Faculdade de F\'{i}sica, Universidade Federal do Pará, Campus Universitário de Tucuruí, CEP: 68464-000, Tucuruí, Pará, Brazil}

     \author{José Tarciso S. S. Junior}
    \email{tarcisojunior17@gmail.com}
\affiliation{Faculdade de F\'{i}sica, Programa de P\'{o}s-Gradua\c{c}\~{a}o em F\'{i}sica, Universidade Federal do Par\'{a}, 66075-110, Bel\'{e}m, Par\'{a}, Brazill}

	\author{Francisco S. N. Lobo} \email{fslobo@ciencias.ulisboa.pt (corresponding author)}
\affiliation{Instituto de Astrof\'{i}sica e Ci\^{e}ncias do Espa\c{c}o, Faculdade de Ci\^{e}ncias da Universidade de Lisboa, Edifício C8, Campo Grande, P-1749-016 Lisbon, Portugal}
\affiliation{Departamento de F\'{i}sica, Faculdade de Ci\^{e}ncias da Universidade de Lisboa, Edif\'{i}cio C8, Campo Grande, P-1749-016 Lisbon, Portugal}

    \author{\\Manuel E. Rodrigues} \email{esialg@gmail.com}
\affiliation{Faculdade de F\'{i}sica, Programa de P\'{o}s-Gradua\c{c}\~{a}o em F\'{i}sica, Universidade Federal do Par\'{a}, 66075-110, Bel\'{e}m, Par\'{a}, Brazill}
\affiliation{Faculdade de Ci\^{e}ncias Exatas e Tecnologia, Universidade Federal do Par\'{a}, Campus Universit\'{a}rio de Abaetetuba, 68440-000, Abaetetuba, Par\'{a}, Brazil}

 \author{Diego Rubiera-Garcia} \email{ drubiera@ucm.es}
\affiliation{Departamento de Física Téorica and IPARCOS, Universidad Complutense de Madrid, E-28040 Madrid, Spain}

     \author{Luís F. Dias da Silva} 
        \email{fc53497@alunos.fc.ul.pt}
\affiliation{Instituto de Astrof\'{i}sica e Ci\^{e}ncias do Espa\c{c}o, Faculdade de Ci\^{e}ncias da Universidade de Lisboa, Edifício C8, Campo Grande, P-1749-016 Lisbon, Portugal}

    \author{Henrique A. Vieira} \email{henriquefisica2017@gmail.com}
\affiliation{Faculdade de F\'{i}sica, Programa de P\'{o}s-Gradua\c{c}\~{a}o em F\'{i}sica, Universidade Federal do Par\'{a}, 66075-110, Bel\'{e}m, Par\'{a}, Brazill}
%%%%%%%%%%%%%%%%%%%%%%%
%%%%%%%%%%%%%%

%%%%%%%%%%%%%%%%%%%%%%%

\begin{abstract}

We consider a recently introduced extension of General Relativity based on the use of the Cotton tensor and dubbed as Cotton gravity, to estimate the size of a new constant $\gamma$ appearing within a spherically symmetric, vacuum solution of such theory. Taking into account its non-asymptotically flat character, we use the inferred size of the central brightness depression of the supermassive object at the heart of the Milky Way galaxy (Sgr A*) by the Event Horizon Telescope to constrain at $2\sigma$) the CG parameter as $\gamma M \approx 3.5 \times 10^{-12}$. We study the potential observational consequences from the smallness of such a value using exact and numerical expressions for the deflection angle, optical images from optically and geometrically thin accretion disks, isoradials, and instability scales (Lyapunov index) of nearly bound geodesics associated to photon rings. Our results point towards the impossibility to distinguish between these two geometries using current and foreseeable techniques in the field of interferometric detection of optical sources.

\end{abstract}

%\pacs{04.70.BW, 04.70.-s}
\date{\today}

\maketitle

%%%%%%%%%%%%%%%%%%%%%%%%%%%%%%%%%%%%%%%%%%%%%%%%%%%%%%%%%%%%%%%%%%%%%%%%%%
\section{Introduction}
%%%%%%%%%%%%%%%%%%%%%%%%%%%%%%%%%%%%%%%%%%%%%%%%%%%%%%%%%%%%%%%%%%%%%%%%%%

Despite the undeniable success of Einstein's General Relativity (GR) in explaining a large number of gravitational phenomena in different regimes \cite{Will:2014kxa}, there are several drawbacks of the theory. On the one hand, the accelerated expansion of the Universe, as found by Type-Ia supernovae observations ~\cite{SupernovaSearchTeam:1998fmf,SupernovaCosmologyProject:1998vns}, can only be accounted for, in the frame of GR, via an hypothetical fluid known as dark energy, whose properties can be assimilated to that of a cosmological constant term yet there is no direct proof of its existence. Similarly, the explanation of galactic rotation curves demands more than 90\% of the total mass of most galaxies to be dominated by a non-baryonic form of matter dubbed as dark matter, which we have been unable to detect in any terrestial observatory yet. On more theoretical grounds, the problems with the unavoidable development of space-time singularities inside black holes and in the early cosmological evolution and the loss of predictability they bring about \cite{Senovilla:2022vlr}, the Hawking process and the associated information loss paradox \cite{Hawking:2005kf}, and the development of a framework unifying gravity and quantum mechanics (quantum gravity \cite{QG}), are challenges that still haunt the gravitational community.

In response to these challenges, a large avenue of research has emerged in the last few decades arranged around alternative approaches to GR, generally dubbed as modified theories of gravity (see e.g. \cite{Capozziello:2011et, Clifton:2011jh} for some reviews). Within this context, recently Harada presented a theory called ``Cotton Gravity" (CG)~\cite{Harada:2021bte}. In this proposal, the effects describing gravity are attributed to the Cotton tensor, and the corresponding field equations contain third-order derivatives. The main interest in this proposal comes from the fact that any solution of the Einstein equations, with or without a cosmological constant term, also satisfies the field equations of CG. Furthermore, the cosmological constant itself emerges as an integration constant out of the field equations. In addition, in \cite{Harada:2022edl} Harada used the post-Newtonian limit of CG to solve the field equations numerically, interpreting the rotation curves of several galaxies without the employ of dark matter sources.

Regarding new solutions within CG, the first non-trivial one with static, spherical symmetry was found in \cite{Harada:2022edl}, which extends the Schwarzschild solution via a new parameter $\gamma$. Such a parameter is incorporated in the vacuum solutions of the theory via the integration of its field equations, and it is interpreted as a measure of the relative strength of the contribution of the Cotton tensor to the space-time geometry. Other non-trivial solutions were found in \cite{Sussman:2023eep,Gogberashvili:2023wed}. On the other hand, in \cite{Mantica:2022flg} it was emphasized that the Codazzi tensor can be derived from the equations of CG, which implies that Harada's formulation can be understood as a modification of the GR equations due to the presence of this residual tensor. This way, with the introduction of the Codazzi tensor, it became possible to express CG through a new formulation, which allowed for the representation  of conformally flat cosmological solutions without a vacuum in CG \cite{Sussman:2023wiw}. 

In order to set constraints in the new parameter $\gamma$ of CG, one can resort to the gravitational lensing of light. This is the deflection of light trajectories as they pass close to massive bodies (stars, black holes, galaxies, etc) and can act like a natural telescope by amplifying the source brightness and even creating multiple images. 
Recent developments related to this effect can be divided into two groups:  microlensing (i.e. the measurement of the collective magnification of various images) and strong lensing \cite{Congdon}. For the sake of this work we are (mostly) interested in the strong field regime, given the fact that it offers the possibility of testing GR beyond the weak-field limit tests carried out so far. In such a limit,  Bozza \cite{Bozza:2002zj} proposed a formalism to calculate the deflection angle for every spherically symmetric metric in terms of a logarithmic expansion. Such a formalism has been very influential in advancing analytical methods to compare theoretical predictions with real images \cite{Bozza:2012by,Pietroni:2022cur}, including alternative spherically symmetric metrics \cite{Tsukamoto:2020iez,Tsukamoto:2020bjm,Tsukamoto:2021fsz,Zhang:2022nnj}.

The black hole shadow and photon ring heuristics are two such examples of visual phenomena caused by strong gravitational lensing \cite{Gralla:2019xty}. The former consists in the dark silhouette of a black hole that is projected in the celestial plane, with a critical boundary separating backwards-traced geodesics that are scattered towards infinity from those that are captured by a black hole \cite{Falcke:1999pj, Cunha:2018acu}. The latter appears as an infinite series of discrete and stacked, exponentially-damped in luminosity, lensed images of the emission surrounding a black hole, under the assumption of an optically thin and non-spherical accretion disk \cite{Gralla:2019drh, Johnson:2019ljv}. The recent progress in high-frequency very-long-baseline interferometry (VLBI) provides a direct way of comparing such heuristics with actual images, by the Event Horizon Telescope (EHT) Collaboration, of the super-heated overflow of the plasma around the supermassive central objects of M87* \cite{EventHorizonTelescope:2019dse} and Sgr A* \cite{EventHorizonTelescope:2022wkp}. In turn, this provides an unprecedented opportunity to perform tests of new gravitational Physics, which has been put to good use very recently, see e.g.  \cite{Vagnozzi:2022moj}.

The main aim of this work is therefore to make use of this opportunity to study the optical appearances of the CG black hole above. In particular, we shall employ the reported bounds on Sgr A*'s shadow by the EHT Collaboration to constrain the CG black hole's parameter space. Next, we generate images of this geometry within this restricted parameter space when illuminated by an optically and geometrically thin accretion disk, modeled by three semi-analytic emission profiles developed by Gralla-Lupsasca-Marrone (GLM) \cite{Gralla:2020srx}, and compare its optical appearance with that of a Schwarzschild black hole. Furthermore, we study the Lyapunov exponents of nearly bound orbits to elaborate on the differences between both geometries. In this work we consider units $G=c=1$.

%%%%%%%%%%%%%%%%%%%%%%%%%%%%%%%%%%%%%%%%%%%%%%%%%%%%%%%%%%%%%%%%%%%%%%%%%%
\section{A spherically symmetric solution of Cotton Gravity}\label{CG}
%%%%%%%%%%%%%%%%%%%%%%%%%%%%%%%%%%%%%%%%%%%%%%%%%%%%%%%

Recently, a generalization of the field equations of GR, dubbed as ``Cotton Gravity" (CG)~\cite{Harada:2021bte}, was proposed. The equations read as follows
\begin{equation}
C_{\alpha\mu\nu}=\kappa^{2}\nabla_{\beta}\Theta_{\phantom{\beta}\alpha\mu\nu}^{\beta},\label{eq_CG}
\end{equation}
where $\kappa^2=8\pi$, is  Newton's constant in suitable units, $\nabla_\beta$ is the usual covariant derivative  built with the Christoffel symbols of the metric, $C_{\alpha\mu\nu}$  is the Cotton tensor defined as
\begin{equation}
C_{\alpha\mu\nu}\equiv\nabla_{\mu}R_{\nu\alpha}-\nabla_{\nu}R_{\alpha\mu}-\frac{1}{6}\left(g_{\alpha\nu}\partial_{\mu}-g_{\alpha\mu}\partial_{\nu}\right)R,
\end{equation}
where $R_{\mu\nu}$ stands for the Ricci tensor and $R$  is the Ricci scalar, and finally the energy-momentum tensor  $\Theta_{\beta\alpha\mu\nu}$ is defined as
\begin{eqnarray}
\Theta_{\beta\alpha\mu\nu} & \equiv & \frac{1}{2}\left(g_{\beta\mu}\Theta_{\alpha\nu}-g_{\alpha\mu}\Theta_{\beta\nu}-g_{\beta\nu}\Theta_{\alpha\mu}+g_{\alpha\nu}\Theta_{\beta\mu}\right) \nonumber \\
&-&\frac{1}{6}\left(g_{\beta\mu}g_{\alpha\nu}-g_{\alpha\mu}g_{\beta\nu}\right)\Theta \,,
\end{eqnarray}
being  $g^{\alpha\nu}\Theta_{\beta\alpha\mu\nu}=\Theta_{\beta\mu}$ and $\Theta=g^{\beta\mu}\Theta_{\beta\mu}$. 

The most remarkable feature of CG gravity is that every solution of the Einstein equations with a cosmological constant term $G_{\mu\nu}+\Lambda g_{\mu\nu}=\kappa^{2}\Theta_{\mu\nu}$ satisfies Eq.\eqref{eq_CG} \cite{Harada:2021bte}, while the opposite is not true. In contrast to GR, however, the cosmological constant in CG is determined by integration of the field equations. Furthermore, the conservation law ($\nabla_{\beta}\Theta_{\phantom{\mu}\nu}^{\beta}=0$) is a natural consequence of the theory, just like in GR, as can be verified by contracting Eq.\eqref{eq_CG} by~$g^{\alpha\nu}$ to yield
\begin{equation}
g^{\alpha\nu}C_{\alpha\mu\nu}=\kappa^{2}\nabla_{\beta}\Theta_{\phantom{\mu}\nu}^{\beta}=0.
\end{equation}

For the sake of this work we shall use the exact vacuum solution obtained by Harada in \cite{Harada:2022edl}. Similarly to GR, where $R_{\mu\nu}=0$, this solution is obtained in Cotton Gravity when $C_{\alpha\mu\nu}=0$ is considered. Such a solution belongs to the static, spherically symmetric class of metrics with line element
\begin{equation}
ds^{2}=-A(r)dt^{2}+\frac{dr^{2}}{A(r)}+C(r)\left(d\theta^{2}+\sin^{2}\theta d\phi^{2}\right).\label{m_cot}
\end{equation}
whose metric function, as found in \cite{Harada:2021bte} reads as 
\begin{equation} \label{Acotton}
A(r)=1-\frac{2M}{r}-\frac{\Lambda}{3}r^{2}+\gamma r \quad   ; \quad C(r)=r^2
\end{equation}
This is the first non-trivial solution found in CG theory, which contains a new parameter $\gamma$, not present in the Schwarzschild case of the GR. As long as its value is non-zero (since for $\gamma=0$ it obviously retrieves the Schwarzschld solution), this means that, even in the case of vanishing cosmological constant, $\Lambda=0$, the space-time is not asymptotically flat\footnote{Such solutions, containing a term that makes the space-time neither to be asymptotically flat nor (anti-)de Sitter are not unknown in the literature, see e.g.  \cite{Mannheim:1988dj, Kiselev:2002dx, Grumiller:2010bz, Ghosh:2015cva, Soroushfar:2015wqa}).}. Given the fact that throughout the paper we shall actually assume $\Lambda=0$, then the presence of the $\gamma r$ term not only drives the asymptotic character of the geometry, but  will also modify the horizons, which are now located as
\begin{equation}
r=\frac{1\pm \sqrt{1-8M \gamma}}{2\gamma}    
\end{equation} 
where $\pm$ for the outer/inner horizons. This is more akin to the global structure of a Reissner-Nordstr\"om solution rather than a Schwarzschild one. For the purposes of this work we are solely interested in the outer (event) horizon, bearing also in mind that light propagation and deflection will be also modified in this theory. Indeed, by comparing the scales at which the gravitational mass and the $\gamma$-parameter become comparable we find a radius
\begin{equation}
 r_h^{\pm} \sim \sqrt{\frac{2M}{\gamma}}   
\end{equation}
for the which $\gamma$-related effects should become relevant.

%%%%%%%%%%%%%%%%%%%%%%%%%%%%%%%%%%%%%%%%%%%%%%%%%%%%%%%%%%%%%%%%%%%%%%%%%%
\section{GEODESICS AND EFFECTIVE POTENTIAL}\label{sec2}
%%%%%%%%%%%%%%%%%%%%%%%%%%%%%%%%%%%%%%%%%%%%%%%%%%%%%%%

\subsection{Photon's motion in Cotton's gravity solution}

In this section, we obtain the null geodesic equation and the effective potential of the black hole described by CG, using the spherically symmetric line element and the specific solution just described above. To this end, we start our analysis by considering the point-particle Lagrangian density given by
\begin{equation}
    {\cal L}=\frac{1}{2}g_{\mu\nu}\dot{x}^{\mu}\dot{x}^{\nu}=\epsilon,\label{Lagran}
\end{equation}
where $\dot{x}^\mu$ is the four-velocity of the particle,  $\epsilon=(-1,0,1)$ indicates the time-like, null, or space-like character of the geodesic, respectively, and a dot denotes the derivative with respect to the affine parameter $\lambda$. The metric~\eqref{m_cot} makes the Lagrangian take the form
\begin{equation}
    {\cal L}=\frac{1}{2}\left[-A\left(r\right)\dot{t}^{2}+\frac{\dot{r}^{2}}{A(r)}+C(r)\dot{\theta}^{2}+C(r) \sin^2 \theta \dot{\phi}^{2}\right]=\epsilon, 
\end{equation}
\par
\begin{figure}[t!]
   \includegraphics[width=\linewidth]{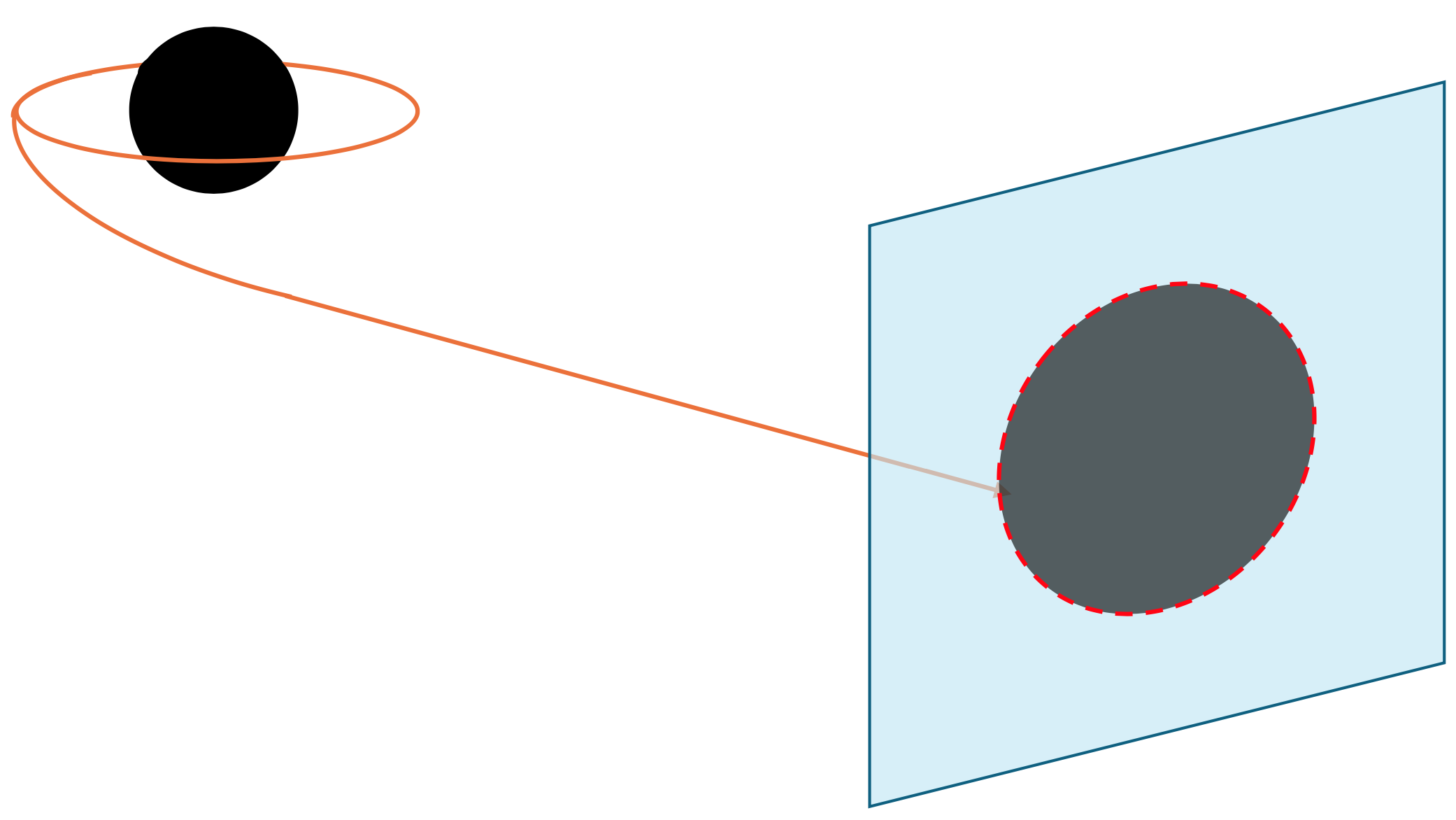}
    \caption{Example of an unstable circular geodesic (orange) in the equatorial plane of a static and spherically symmetric black hole (black circle). In the celestial plane (blue), the critical curve (dashed red) is comprised of light rays which asymptotically approach the photon sphere.}
    \label{fig:BHappearance}
\end{figure}

Frow now on we restrict our analysis to null geodesics, $\epsilon=0$. Furthermore, given the spherically symmetric character of the space-time, we can consider the motion in the equatorial plane, $\theta=\pi/2$, without any loss of generality. In addition, since the Lagrangian~\eqref{Lagran} does not directly depend on $t$ and $\phi$, the Euler-Lagrange equations imply the conservation of two quantities, $E=-A\left(r\right)\dot{t},
L=C\left(r\right)\dot{\phi},$ which are the photon's energy and angular momentum, respectively. This way, Eq.(\ref{Lagran}) with the conserved quantities above leads to
\begin{equation}
\dot{r}^{2}=\frac{1}{b^2}-V_{eff}(r). \label{eq_geo} 
\end{equation}
where we have conveniently re-absorbed a factor $L$ in the affine parameter $\lambda$, and introduced the impact parameter $b \equiv L/E$. In this equation, the effective potential reads as
\begin{equation}
    V_{eff}(r)=\frac{A(r)}{C(r)} =\frac{1}{r^2}-\frac{2 M}{r^3}+\frac{\gamma}{r} \label{V}
\end{equation}
where the last equality correspond to the metric function considered in this work, Eq.\eqref{Acotton}.
The maximum of the effective potential determines the region of unstable orbits, dubbed as the \textit{photon sphere}, whose radius is denoted by $r=r_p$, and the corresponding impact parameter as the \textit{critical impact parameter}, dubbed by $b_c$ (see e.g. Fig. \ref{fig:BHappearance}).
The latter separates those orbits that are captured by the black hole, $b<b_c$, from those that are scattered back to asymptotic infinity, $b>b_c$. Those that have $b=b_c$ may linger indefinitely at the photon sphere, turning several times around the black hole. To determine these quantities, we note that circular light orbits obey $\dot{r}=0$ and $\ddot{r}=0$ simultaneously. Under these conditions, $r_p$ and $b_c$ are expressed for the metric function~\eqref{Acotton} as
\begin{subequations}
\begin{align}
    r_p&=\frac{\sqrt{1+6 \gamma  M}-1}{ \gamma }\label{eq:rps};\\
    b_c&=\sqrt{\frac{C(r_p)}{r_p}}=\frac{3 \sqrt{6}}{\sqrt{\frac{\left(1+ \sqrt{1+6 \gamma   M}+6 \gamma   M \sqrt{1+6 \gamma  M}+9 \gamma  M\right)}{M^2}}}.\label{bc}
\end{align}
\end{subequations}
which recovers the critical impact parameter $b_c=3\sqrt{3}M$ and the photon sphere radius $r_p=3M$ of the Schwarzschild geometry, in the $\gamma \rightarrow 0$ limit.
The geodesic equation (\ref{V}) can be more conveniently written, with the help of the conserved quantities, as 
\begin{equation}
    \left(\frac{dr}{d\phi}\right)^{2}=C(r)\left[\frac{1}{b^{2}}-V_{eff}\right].\label{eqphi}
\end{equation}

Photons whose impact parameter lie within the region $b<b_c$ eventually meet the event horizon. This region, bounded by $b=b_c$ (dubbed as the critical curve and corresponding to the projection, on the observer's screen, of the photon sphere), is marked by a strong decrease in the observed luminosity, as projected onto the celestial plane of a distant observer. This is conventionally known in the literature as the {\it shadow}. To calculate its angular size we shall adopt the formalism described by \cite{Perlick:2021aok}. In their approach, they assume a static observer located at ($r_0$, $\theta_0=\pi/2$) and projecting an angle $\theta$ with respect to the radial coordinate $r$ given by
\begin{equation}    \tan\theta=\sqrt{\frac{C(r)}{A(r)}}\frac{d\phi}{dr}\Bigg|_{r=r_{0}}.\label{tan}
\end{equation}
The shadow of the black hole, understood as the region which borders (on the observer's plane image) the critical curve, is thus constructed by the light rays emitted from the observer back to the source.

By introducing the definition
\begin{equation}
    h^{2}(r)=\frac{C(r)}{A(r)},\label{h}
\end{equation}
in terms of which Eq.(\ref{eqphi}) can be written as
\begin{equation}
    \left(\frac{dr}{d\phi}\right)^{2}=\frac{C(r)}{A(r)}\left(\frac{h^{2}(r)}{h^{2}(r_p)}-1\right).\label{drdphi}
\end{equation}
then by substituting into into Eq.\eqref{tan} we find  the expression
\begin{equation}
    sin^{2}\theta=\frac{h^{2}(r_p)}{h^{2}(r_{0})},
\end{equation}
which yields the angular size of the black hole's shadow $\theta_s$ as
\begin{equation}
\sin^{2}\theta_s=\frac{b_{c}^{2}}{h^{2}(r_{0})}=\frac{b_{c}^{2}A(r_0)}{C(r_{0})}.
\end{equation}
For the CG solution, this angular shadow's size reads as 
\begin{widetext}
\[
\sin^{2}\theta_{s}=\frac{2}{r_{0}^{2}\left(\gamma +8\gamma^{2}M\right)}\left(-\frac{1}{\gamma}+\frac{6M\sqrt{\left(1+6\gamma M\right)}}{1}+\frac{\sqrt{\left(1+6\gamma M\right)}}{\gamma}-9M\right)\left(-\frac{2M}{r_{0}}+\gamma r_{0}+1\right)\,.
\]
\end{widetext}

\begin{figure}[t!]
    \centering
   \includegraphics[width=0.95\linewidth]{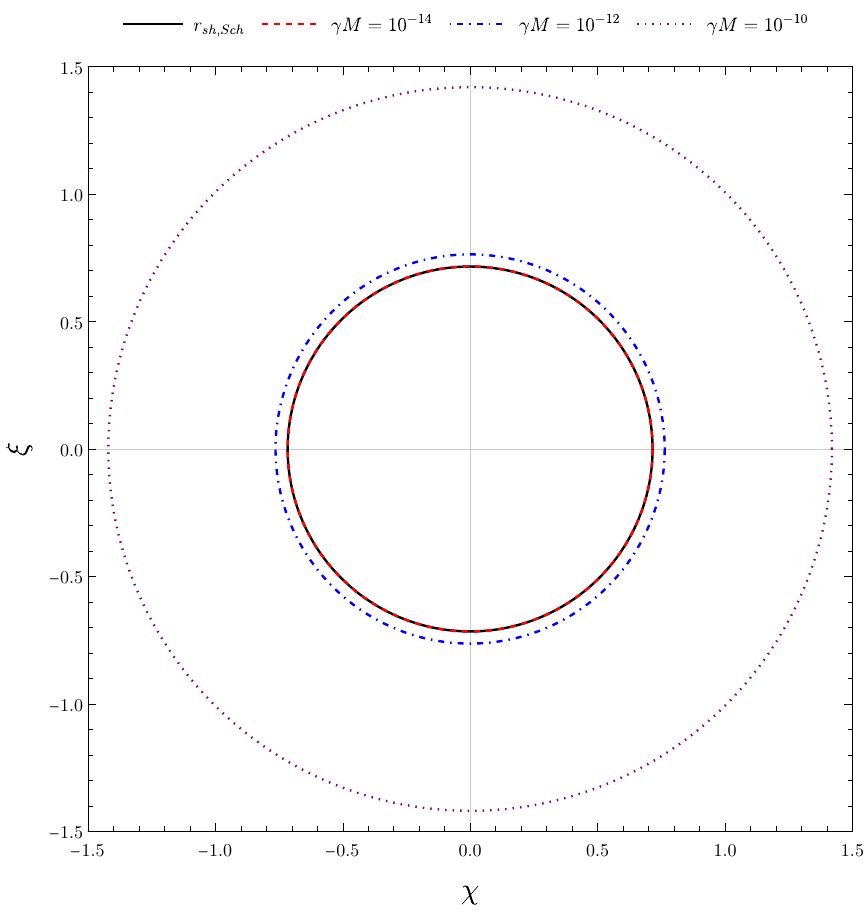}
    \caption{Black hole shadow $r_{sh}$ boundary as seen in the celestial plane $(\chi,\xi)$ of an equatorial observer, for the Schwarzschild geometry (black) and for CG with $\gamma M \in \{10^{-14},10^{-12},10^{-10}\}$ (colored and dashed lines). The celestial coordinates are computed considering the logarithm of the shadow radius (i.e. $\chi=\text{Log}_{10}\left(\frac{r_{sh}}{M}\right) \cos(\phi)$; $\xi=\text{Log}_{10}\left(\frac{r_{sh}}{M}\right) \sin(\phi)$).}
    \label{shadow}
\end{figure}

If, instead, we are interested in the shadow's radius as seen by a far-away observer, for asymptotically flat space-times one can approximate $\sin(\theta) \sim \theta$ and $\theta_s=b_c/r_0$ so the shadow's is simply the critical impact parameter (in units of length), that is, $r_{sh}=b_c=r_p/g_{tt}(r_p)^{1/2}$. However, in our non-asymptotically flat space-time, we need to specify the metric functions at the specific observer radius as \cite{Perlick:2021aok}
\begin{equation}
r_{s}=r_{p}\sqrt{\frac{g_{tt}(r_{0})}{g_{tt}(r_{p})}}.\label{r_shadow2}
\end{equation}
which for CG  reads as
\begin{equation} \label{rs_CG}
r_s=\frac{\left(\sqrt{a}-1\right)}{\gamma}\sqrt{\frac{\left(\sqrt{a}-1\right)\Bigl[r_{0}\left(\gamma r_{0}+1\right)-2M\Bigr]}{r_{0}\left(1-\sqrt{a}+4\gamma M\right)}}\,.
\end{equation}
where $a=1+6\gamma M$. In Fig. \ref{shadow} we depict the behaviour of the shadow's radius in CG with the parameter $\gamma$, considering the mass and distance estimates corresponding to  Sgr A$^*$, reported in Table \ref{tab:my-table}.

\subsection{Constraints to the $\gamma$ parameter from shadow's radius}

We now aim to restrict the parameter space of $\gamma$ by assessing the compatibility of the CG geometry's corresponding shadow radius $r_s$ with the EHT's inferred shadow size of Sgr A* using the data above. This is so because, as argued in \cite{EventHorizonTelescope:2022xqj}, it is reported that the observed shadow size of Sgr A* can be inferred by utilizing the size of the bright emission ring as its proxy, subject to a calibration factor that multiplicatively accounts for the theoretical and observational uncertainties. Such a correlation becomes possible thanks to priors on Sgr A*'s mass-to-distance ratio obtained via orbital dynamics measurements of the stars closest to the galactic center (i.e. dubbed S-stars), and which are reported in Table \ref{tab:my-table}.  These measurements allow the EHT collaboration to estimate the fractional deviation $\delta$ between Sgr A*'s inferred shadow size $r_h$ and that of a Schwarzschild black hole $r_{sh,Sch}= 3 \sqrt{3} M$, with angular radius $\omega_g = M/D$  via the formula
\begin{equation}\label{eq:fractionaldeviation}
    r_{sh}/M=(\delta +1) 3 \sqrt{3} \ ,
\end{equation}
which, under the assumption of a normal distribution of the estimation uncertainties, yields the following shadow size bounds \cite{Vagnozzi:2022moj}
\begin{equation} \label{eq:1sigma}
    4.55 \lesssim r_{sh}/M \lesssim 5.22 \ ,
\end{equation}
at $1\sigma$ deviation and
\begin{equation} \label{eq:2sigma}
    4.21 \lesssim r_{sh}/M \lesssim 5.56 \ ,
\end{equation}
at $2\sigma$ deviation. It is worth mentioning that many different geometries have been constrained in the literature following this observation, see e.g. \cite{EventHorizonTelescope:2021dqv,Ghosh:2022kit,Afrin:2021wlj,Afrin:2022ztr,Afrin:2021imp} and the review \cite{Vagnozzi:2022moj}.

\begin{table}[t!]
\centering
\resizebox{\linewidth}{!}{%
\def\arraystretch{1.4}
\begin{tabular}{cccc}
\hline \hline
\multicolumn{4}{c}{Parameter values}                                                                               \\ \hline
\hspace{1 mm}Survey\hspace{1 mm} & \hspace{1 mm}$M (\times 10^6 M_{\odot})$\hspace{1 mm} & \hspace{1 mm} $D$\hspace{1 mm} (kpc) & \hspace{1 mm}Reference\hspace{1 mm} \\ 
Keck                             & $3.951 \pm 0.047$                                     & $7.953 \pm 0.050 \pm 0.032$         & {\cite{Do:2019txf}} \\ \hline
\end{tabular}%
}
\caption{Sgr A* Mass and distance.}
\label{tab:my-table}
\end{table}

We note that, to quantify the theoretical biases in the calibration factor utilized in the inference of the shadow's size of Sgr* A, the EHT collaboration performs simulations of black hole images to study a wide range of effects concerning the accretion flow properties and the underlying space-time. In this regard, the collaboration assumes the Kerr metric by default, although they also cover black hole metrics that either deviate from the Kerr space-time or describe alternative geometries, such as wormholes or boson stars. Because their analysis implicitly assumes asymptotically flat, GR-based solutions, this does not include the Cotton geometry among other theories than stray too much from GR description of gravitational phenomena, by performing tests with the above constraints we introduce a theoretical bias in our own analysis, since we assume these results to be valid as means to set constraints for the CG spacetime. The extent to which the Cotton geometry could modify the above fractional deviation, and the constraints stemming from it, could be evaluated by performing a similar analysis to that by the EHT collaboration. However, such an effort lies well beyond the purpose of this work. 

\begin{figure}[t!]
   \includegraphics[width=\linewidth]{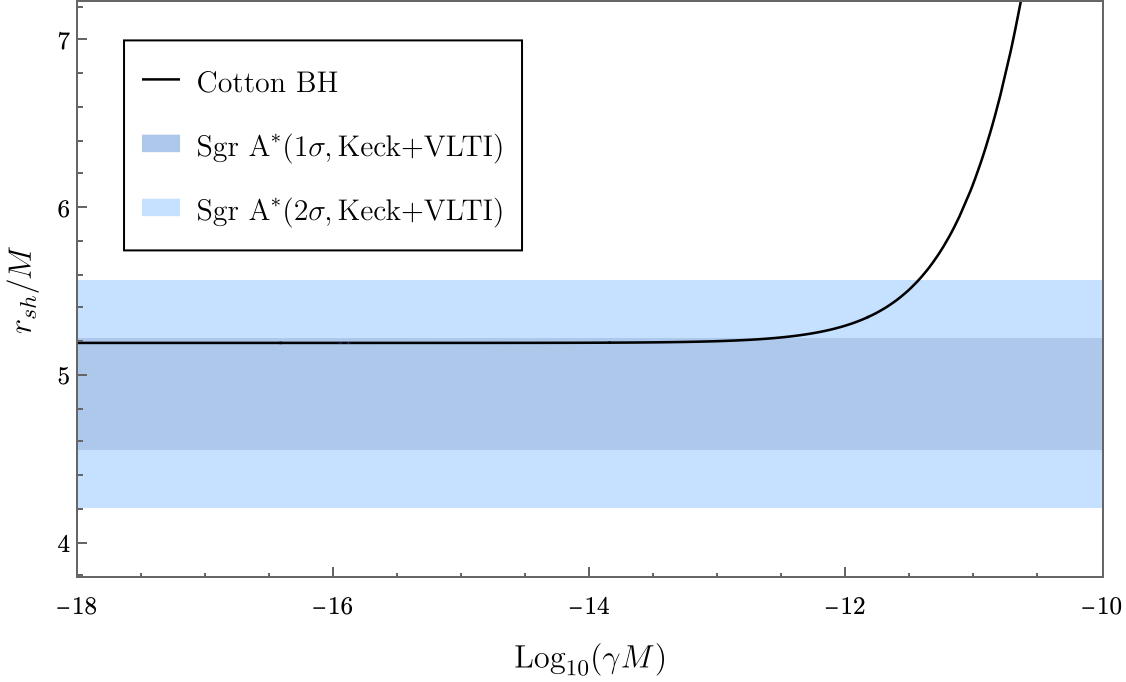}
    \caption{Shadow radius $r_{sh}$ of a Cotton Gravity black hole (solid black curve) with a metric function provided by Eq.\eqref{Acotton}, as a function of the logarithm of $\gamma M$, and calculated according to equation \eqref{rs_CG}, for an observer located at $r_O \sim 8000 \  \text{kpc}$. The shaded blue and light blue regions represent the $1\sigma$ and $2\sigma$ EHT constraints, respectively, to the shadow radius of Sgr A*.}
    \label{fig:shadowconstraints}
\end{figure}

To perform tests on the CG metric with its shadow radius, we compare the evolution of $r_s$ as a function of the logarithm of $\gamma M$, in Fig. \ref{fig:shadowconstraints}. We restrict our analysis to positive values of $\gamma$, which are required to describe the large-distance gravitational acceleration. 
Because we are dealing with a non-asymptotically flat space-time, one must take into account that its asymptotic Arnowitt-Deser-Misner (ADM) mass \cite{Arnowitt:1962hi} is undefined for an observer at radial infinity and cannot be associated with the mass parameter appearing in the metric.
To circumvent this issue, we shall take a similar approach to \cite{Vagnozzi:2022moj} and report the the shadow radius in units of the local mass $M$, based on the reasoning that the stellar orbital dynamics of the S group of stars probe this quantity, since they are influenced by the local gravity of Sgr A*. 
Additionally, since the CG black hole's shadow radius depends explicitly on the observer's radial position, recall Eq.(\ref{rs_CG}), we compute the shadow radius considering an observer located at $r_O \sim 8000 \  \text{kpc}$ \cite{Do:2019txf}. 

Upon inspection of Fig. \ref{fig:shadowconstraints}, we verify that $r_{sh}$ starts to increase around values of $\gamma M \sim 10^{-12}$. Below this value, the CG black hole shadow radius coincides with that of a Schwarzschild black hole. We note that this is qualitatively the same behavior that is evidenced in \cite{Vagnozzi:2022moj} for the Rindler metric. Actually, the contribution of $\gamma$ to the shadow radius becomes significant when the term $\gamma r \gg 1$, while the upper $2\sigma$ bound in Eq.(\ref{eq:2sigma}) becomes saturated at a value of approximately $\gamma M \simeq 3.525 \times 10^{-12}$.
For the values of $\gamma M$ predicted by galaxy radial acceleration rotation curves in \cite{Harada:2022edl, Gogberashvili:2023wed}, or the recent constraints obtained by analyzing CG's influence on CMB power spectrum and large scale structures \cite{Xia:2024tps} (i.e. both estimating $\gamma M \sim 10^{-17}$), the shadow radius of the Cotton black hole lies well within the constraints imposed by Sgr A*'s shadow bounds, though there is no appreciable difference between it and its Schwarzschild counterpart. This result is not surprising, given that $\gamma$ encodes large-distance gravitational effects within the CG framework. 
Beyond the $\gamma$ parameter range considered in Fig. \ref{fig:shadowconstraints}, we also found that the Cotton black hole shadow radius remains within the $2\sigma$ bounds for $\gamma M \sim 10^{20}$. Such values describe extremely large and unphysical gravitational acceleration effects, exceeding any other current estimates by several orders of magnitude. This merited their exclusion from analysis in the following sections. \\

%%%%%%%%%%%%%%%%%%%%%%%%%%%%%%%%%%%%%%%%%%%%%%%%%%%%%%%%%%%%%%%%%%%%%%%%%%%%%%%%%%%%%%%%%%%%%%%%%%%%%%%%%%%%%%%%%%%%%%%%%%%%

%%%%%%%%%%%%%%%%%%%%%%%%%%%%%%%%%%%%%%%%%%%%%%%%%%%%%%%%%%%%%%%%%%%%%%%%%%%%%%%%%%%%%%%%%%%%%%%%%%%%%%%%%%%%%%%%%%%%%%%%%%%%
\section{Gravitational lensing \label{sec:lens}}
%%%%%%%%%%%%%%%%%%%%%%%%%%%%%%%%%%%%%%%%%%%%%%%%%%%%%%%%%%%%%%%%%%%%%%%%%%%%%%%%%%%%%%%%%%%%%%%%%%%%%%%%%%%%%%%%%%%%%%%%%%%%
This section deals with the phenomenon of gravitational lensing for CG. First, we will obtain the exact analytical expression for the deflection angle in terms of elliptic integrals. Then we will calculate an approximate expression valid for the weak field regime.

\subsection{Exact deflection angle \label{sec:exactlens}}

To proceed with our derivation, it is convenient to make the following variable change in the radial coordinate $u = 1/r$ in the expression~\eqref{eqphi}. This allows us to re-write the photon equation of motion as
\begin{equation}
    \left( \frac{du}{d \phi} \right)^2 = 2 M G(u),
\end{equation}
where the function $G(u)$ reads, for CG, as
\begin{equation}
    G(u) =u^3 - \frac{u^2}{2M}  - \frac{u \gamma}{2M} + \frac{1}{2Mb^2}.
\end{equation}

The general expression for the deflection angle at a finite distance has been worked out in \cite{Ishihara:2016vdc} as
\begin{equation}
    \alpha = \sqrt{\frac{1}{2M}} \left( \int_{u_R}^{u_0} \frac{du}{\sqrt{G(u)}}+\int_{u_S}^{u_0} \frac{du}{\sqrt{G(u)}} \right) + \Psi_R - \Psi_S.
    \label{eq:alphageral}
\end{equation}
Where $u_0 = 1/r_0$ is the inverse of the distance of closest approach, while $u_R = 1/r_R$ and $u_S$ are the inverse of the radius of the observer and the source, respectively. The function $\Psi$ encodes a correction term accounting for the finite distance, and is given by
\begin{equation}
    \Psi = \arcsin \bigl[ bu \sqrt{A(u)} \bigr],
\end{equation}
where the indices refer to the position of the observer (R) and the source (S). Note that if we take the limit of  $\gamma \rightarrow 0 $, and consider also that observer and source are located at infinity we will have
\begin{equation}
    \alpha_{SC} = \sqrt{\frac{2}{M}}\int_{0}^{u_0} \frac{du}{\sqrt{G(u)}} - \pi,
\end{equation}
which is the well-known deflection angle formula for the Schwarzschild solution.  This change in the definition of the deflection angle is necessary because the Cotton metric is not asymptotically flat. Moving on, the integrals of equation \eqref{eq:alphageral} can be solved by rewriting the function $G(u)$ as
\begin{equation}
    G(u) = (u-u_1)(u-u_2)(u-u_3),
\end{equation}
where $u_1,u_2$ and $u_3$ are the roots of the cubic polynomial. 
Since $u_0=1/r_0$ is one of the roots of this polynomial, we have chosen (for reasons that will become clear later) $u_2 = 1/r_0$. Now we can write
\begin{eqnarray}
   &&  u^3 -u^2(u_1 +u_3 +1/r_0) +u (u_1u_3 +u_1/r_0 +u_3/r_0) 
   \nonumber \\
   &&
   - u_1u_3/r_0  = u^3 - \frac{u^2}{2M} - \frac{u \gamma}{2M} + \frac{1}{2Mb^2}.
    \label{eq:solvegu}
\end{eqnarray}
After comparing the coefficients in Eq.\eqref{eq:solvegu}, we get
\begin{equation}
    \begin{aligned}
        & u_1 +u_3 +1/r_0 =   \frac{1}{2M}, \\
        & u_1u_3 +u_1/r_0 +u_3/r_0 =  - \frac{ \gamma}{2M}.
    \end{aligned}
\end{equation}
Solving the above system we find that $u_1$ and $u_3$ are  given by the following expressions
\begin{equation}
    \begin{aligned}
        &u_1 = \frac{-2 M
   +r_0 \mp \sqrt{-(12 M^2 -8 \gamma  M r_0^2-4 M r_0-r_0^2)}}{4 M r_0} , \\
        &u_3 = \frac{-2 M+r_0 \pm \sqrt{- \left(12 M^2-8 \gamma  M r_0^2-4 M
   r_0-r_0^2\right)}}{4 M r_0},
    \end{aligned}
     \label{eq:u1u3}
\end{equation}
whose signs cannot be chosen independently.
Note that we have two different solutions, one of which is $u_1> u_3$ and another that $u_3> u_1$.
Since our goal is that these roots fulfill the condition $u_3> u_2> u_1$, we choose the negative sign in the expression $u_1$ and the positive sign in the expression $u_3$. Now both integrals of \eqref{eq:alphageral} can be put into the form
\begin{equation}
    I_1(u^{\prime}) =  \int_{u^{\prime}}^{u_2}  \frac{du}{\sqrt{(u-u_1)(u-u_2)(u-u_3)}}
\end{equation}
with $u_3>u_2>u^{\prime}>u_3$ and $u^{\prime}$ being any finite distance. The result of the above integral is %\cite{}
\begin{equation}
    I(u^{\prime}) = \frac{2}{\sqrt{u_3-u_1}} F(\delta,k^2),
\end{equation}
where $F(\delta,k^2) $ is a incomplete first-order elliptic integral of the parameters
\begin{eqnarray}
\delta &=& \arcsin{\sqrt{\frac{(u_3-u_1)(u_2-u^{\prime})}{(u_2-u_1)(u_3-u^{\prime})}}}, \\
    k^2 &=& \frac{u_2-u_1}{u_3-u_1}.
\end{eqnarray}

 To get the result of the integrals given in \eqref{eq:alphageral} we just need evaluate $I(u_R)+I(u_S)$. The complete result will also depend on the correction given by the functions $\Psi(u_R) - \Psi(u_S)$. 
For the Cotton metric, the function $\Psi$ is 
\begin{equation}
    \Psi(u) = \arcsin\left(b u \sqrt{-2 M u+\frac{\gamma }{u}+1}\right).
    \label{eq:Psi}
\end{equation}
while we have for the parameters
\begin{widetext}
\begin{equation}
    \delta(u^{\prime}) = \arcsin\Biggl[ 2 \sqrt{2} \Biggl( - \left(M r_0 (u^{\prime} r_0-1) \sqrt{r_0^2 c}\right) \times \Biggl( \Bigl(\sqrt{r_0^2 c}  +6 M  r_0- r_0^2\Bigr)
   \left(\sqrt{ r_0^2 c}-2 M
    r_0 (2 R  r_0+1)+ r_0^2\right)\Biggr)^{-1} \Biggr)^{1/2}    \Biggr],
    \label{eq:delta}
\end{equation}
\end{widetext}
where $c=-12 M^2+4 M
   r_0 (2 \gamma  r_0+1)+r_0^2$
\begin{equation}
   k^2= \frac{\sqrt{ r_0^2-12 M^2+4 M r_0 (2 \gamma 
  r_0+1)}+6 M -r_0}{2\sqrt{r_0^2 -12 M^2+4 M
   r_0 (2 \gamma  r_0+1)}}.
   \label{eq:k}
\end{equation}
which closes all the expressions we needed to characterize the deflection angle in CG. It is also interesting to re-write these expressions in terms of the impact parameter by inverting its definition, in CG, as 
\begin{widetext}\label{eq:r0eb}
\[
r_0 = \frac{b^2 \gamma }{3}+\frac{2}{3} b \sqrt{b^2 \gamma ^2+3} \cos \left(\frac{1}{3} \cos
   ^{-1}\left(\frac{\sqrt{-\frac{1}{b^4 \left(-\gamma ^2\right)-3 b^2}} \left(-2 b^6 \gamma ^3-9 b^4 \gamma
   +54 b^2 M\right)}{2 \left(b^4 \left(-\gamma ^2\right)-3 b^2\right)}\right)\right).
\]
\end{widetext}

\begin{figure}[t!]
    \centering
   \includegraphics[width=\linewidth]{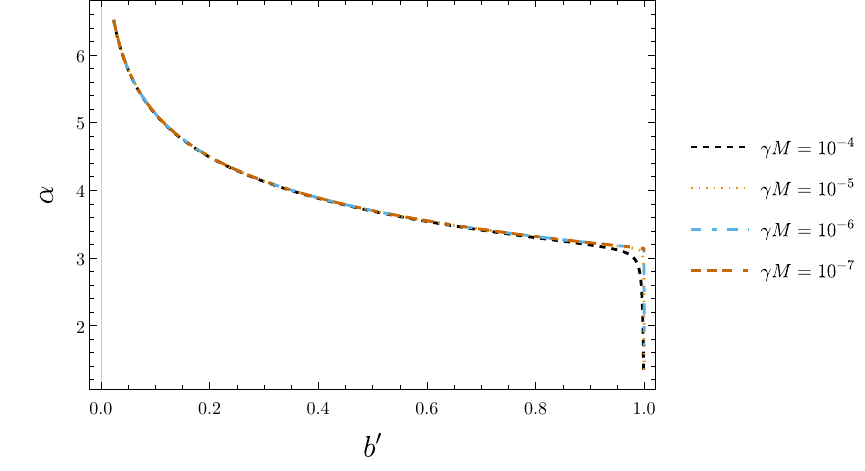}
    \caption{Graphical representation of the deflection angle $\alpha$ (in radians) as a function of the reduced impact parameter $b^{\prime}$ for values of the CG parameter $\gamma M= \{10^{-4},10^{-5},10^{-6},10^{-7}\}$.}
    \label{fig:alphagamma1}
\end{figure}

\begin{figure}[t!]
    \centering
   \includegraphics[width=\linewidth]{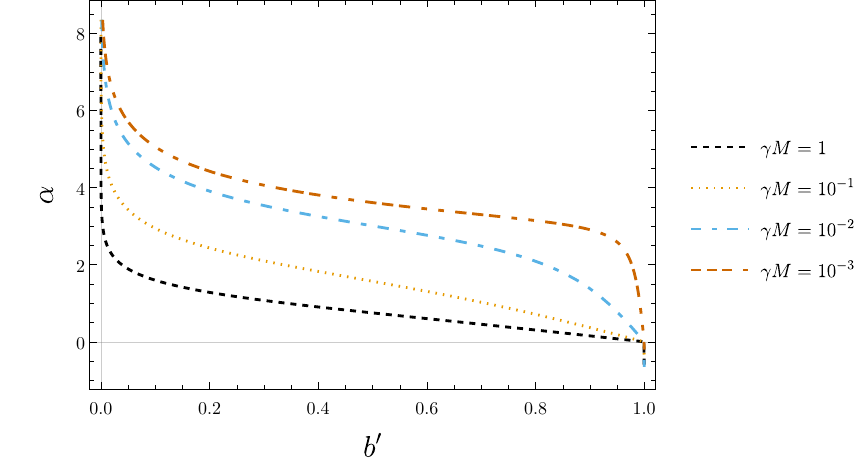}
    \caption{Graphical representation of the deflection angle $\alpha$ (in radians) as a function of the reduced impact parameter $b^{\prime}$ for values of the CG parameter $\gamma M= \{1,10^{-1},10^{-2},10^{-3}\}$.}
    \label{fig:alphagamma2}
\end{figure}

For the sake of the graphical representation of the deflection angle we prefer to use the variable $b^{\prime} = 1-b_c/b$. This variable is limited to a value between $0 \leq b^{\prime} \leq 1$, where zero represents the strong field limit and one the weak field one. In Figs. \ref{fig:alphagamma1} and \ref{fig:alphagamma2} we show the deflection angle as a function of the variable $b^{\prime}$ for values of $\gamma$M in the range from $10^{-7}$ to $1$. We see that $\alpha$ decreases with increasing $\gamma$ and that there is no significant change in the values of $\alpha$ from $\gamma M<10^{-4}$. If we increase the value of the parameter $\gamma M$  beyond the order of magnitude of $10^{-4}$, the deflection angle begins to give negative values. Observationally, this value diverges for $b^{\prime} =0$ and tends to zero for $b^{\prime} =1$. We expect our theoretical result to reproduce this behaviour.
In other words, for the phenomenon of light deflection, Cotton's parameter is limited to values smaller than $\gamma M = 10^{-4}$.

%%%%%%%%%%%%%%%%%%%%%%%%%%%%%%%%%%%%%%%%%%%%%%%%%%%%%%%%%%%%%%%%%%%%%%%%%%

\section{Images of black holes with thin accretion disks\label{sec:images}}
%%%%%%%%%%%%%%%%%%%%%%%%%%%%%%%%%%%%%%%%%%%%%%%%%%%%%%%%%%%%%%%%%%%%%%%%%%%%%%%%%%%%%%%%%%%%%%%%%%%%%%%%%%%%%%%%%%%%%%%%%%%%

Having restricted the $\gamma$-space parameter, we now proceed to generate the images of the Cotton black hole (by pushing the value saturating the upper $2\sigma$ bound of Sgr A* shadow) and qualitatively compare its appearance to that of a Schwarzschild black hole of the same mass parameter. We shall consider the setting of a black hole illuminated by an optically (i.e. fully transparent to its own radiation) and geometrically thin accretion disk, when observed at both a face-on inclination (i.e. $\theta_o = 0^{\circ}$) and at a nearly edge-on inclination (i.e. $\theta_o = 80^{\circ}$).  

\begin{figure*}[t!]
\centering
\subfloat{
\includegraphics[width=7.35cm]{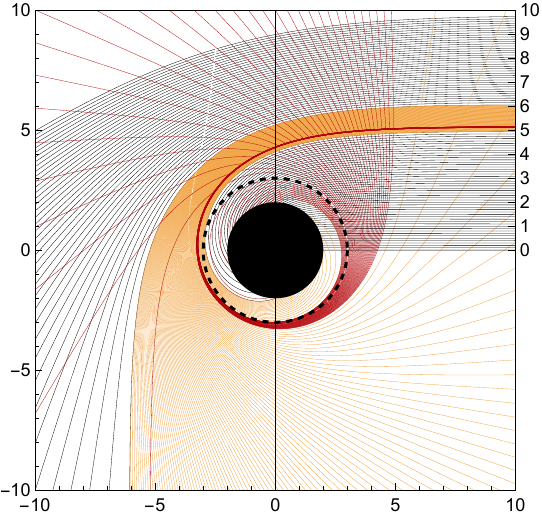}
\includegraphics[width=7.35cm]{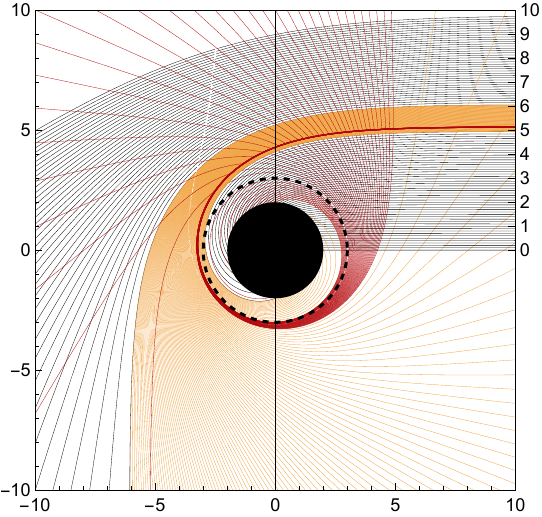}
}\quad
\caption{Null geodesics in the Schwarzschild space-time (left panel) and in the CG space-time ($\gamma M = 3.525 \times 10^{-12}$) as a function of impact parameter $b$ normalized with respect to the mass parameter $M$. The photon trajectories are categorized according to the number of half-turns $n$  around the black hole as the ($n=0$) direct image (black), the $n=1$ photon ring (gold), and the $n=2$ photon ring (red), and with an impact parameter spacing of $1/5$, $1/50$, and $1/500$, respectively. The black circle depicts the inner region to the event horizon. The dashed black circumference represents the photon sphere radius. The black hole equatorial plane is represented by the black vertical line. The observer's screen is located on the right-side axis.}
\label{fig:BHgeodesics}
\end{figure*}

We begin our considerations from Eq.\eqref{eqphi}, which we conveniently re-write under the form
\begin{equation}
    \frac{d\phi}{dr}=\pm \frac{b}{C(r)} \sqrt{ \frac{1}{1-b^2V_{eff}}},
\end{equation}
describing the variation in a photon's azimuthal angle as a function of its radial coordinate. By numerically ray-tracing the previous equation for a bunch of trajectories within some range of the impact parameter $b$, we compute the null geodesics from an observer's screen backwards towards their point of origin. Fig. \ref{fig:BHgeodesics} displays the geodesics in the Schwarzschild space-time (left panel) for $M=1$, and in the CG space-time (right panel), the latter with a value of $\gamma M = 3.525 \times 10^{-12}$. Upon inspection of Fig. \ref{fig:BHgeodesics}, one highlights the absence of any significant qualitative differences between the strong lensing properties of the two geometries, under the considered parameter values. In addition to larger distances, the effects of $\gamma$ become more noticeable at much higher orders of magnitude than those considered here, as is discussed in section \ref{sec:IsoR} below.

To simulate the emission from an accretion disk we resort to the semi-analytic models provided by Gralla, Lupsasca and Marrone in Ref. \cite{Gralla:2020srx}, which describe the radial emission profile of an accretion disk based on a Standard Unbound (SU) Johnson distribution function given by
\begin{equation}\label{eq:GLMfunction}
    I(r;\gamma,\mu,\sigma)=\frac{e^{-\frac{1}{2}\left[\beta+\text{arcsinh}(\frac{r-\mu}{\sigma})\right]^2}}{\sqrt{(r-\mu)^2+\sigma^2}},
\end{equation}
where the $\beta, \mu$ and $\sigma$ parameters modify the skewness, peak location and kurtosis of the distribution, respectively. Such models are designed to reproduce the most prominent image features of time-averaged General Relativistic Magneto-Hydrodynamic (GRMHD) simulations of a thick accretion flow around a Kerr black hole, providing a less computationally taxing alternative towards the generation of black hole images \cite{Gralla:2020srx}. Given the fact that such parameters are also indirectly dependent on the background geometry (for instance, via the location of its event horizon or photon spheres), they can be also employed to generate simulated images of alternative black holes both within and beyond GR, this way producing templates of expected optical appearances under different assumptions for the accretion flow.

The SU models describe an infinitesimally thin, axisymmetric, radially dependent emission profile, assumed to be monochromatic in the disk's frame of reference and optically thin at the emitting frequency. In the absence of absorption, the relativistic radiative transfer equation \cite{Gold:2020iql} implies that the specific intensity $I_{\nu}/\nu^3$ is conserved along a light ray such that $I_{\nu_o}=g^3 I_{\nu_e}$, where the subscripts $o$ and $e$ denote the observer and emitter frames, respectively. The quantity $g$ is defined as the redshift factor $g=\nu_o/\nu_e = \sqrt{A(r)/A(r_{\infty})}$. Since the disk is geometrically thin, a light ray gains additional brightness for each time it intersects the accretion disk while on its way to a distant observer (i.e. every $n$ half-orbit). Thus, considering a monochromatic emitter $I_{\nu_e}\equiv I(r)$, the integrated (bolometric) intensity $I=\int I_{\nu_o} d \nu_o$ in the observer's plane is computed according to
\begin{equation}\label{eq:integratedIntensity}
    I_{o}(r)=\sum_{n}\left(\frac{A(r)}{A(r_o)}\right)^2\left.I(r)\right|_{r=r_n}.
\end{equation}
%I_{o}(r)=\sum_{n} \left. \left(\frac{A(r)}{A(r_o)}\right)^2 I(r) \right|_{r=r_n} .

\begin{figure}[h!]
    \centering
   \includegraphics[width=\linewidth]{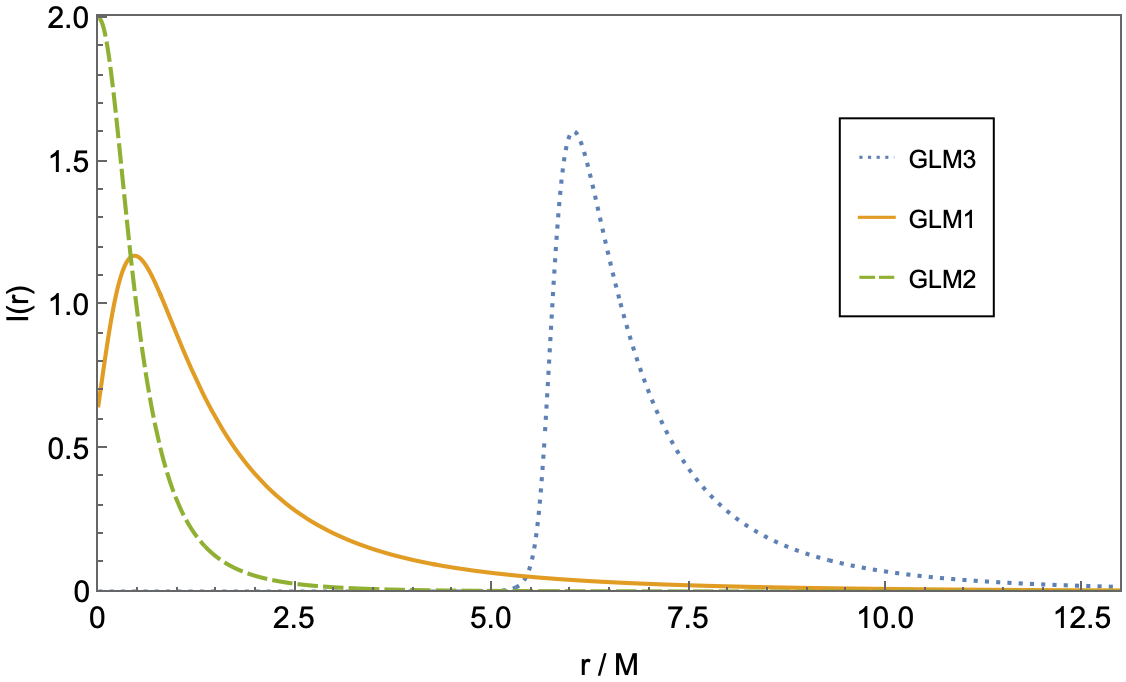}
    \caption{Emission profiles corresponding to the GLM1, GLM2 and GLM3 models described by the choice of parameters in \eqref{eq:GLM1}, \eqref{eq:GLM2} and \eqref{eq:GLM3}, respectively, of the intensity function (\ref{eq:GLMfunction}).}
    \label{fig:GLMmodels}
\end{figure}

For the purpose of generating black hole images we will restrict out analysis up to the $n=2$ photon ring, as higher-order photon rings will provide a negligible contribution to the net luminosity (as shall be explained in the next section). We shall suitably adapt the three models introduced in \cite{Gralla:2020srx} to a spherically symmetric scenario, to simulate the accretion flow around the Schwarzschild and Cotton black holes, with the parameters
\begin{eqnarray}
\text{GLM1}&:&   \beta=-\frac{3}{2} \, , \quad \mu=0 \, , \qquad \  \sigma=\frac{M}{2} \label{eq:GLM1} \ , \\
\text{GLM2}&:&   \beta=0 \, , \qquad \mu=0 \, , \qquad \  \sigma=\frac{M}{2} \    \label{eq:GLM2}, \\
\text{GLM3}&:& \beta=-2 \,, \quad \  \mu=\frac{17M}{3} \, , \    \sigma=\frac{M}{4} \label{eq:GLM3}  \ .
\end{eqnarray}
whose corresponding emission profile is depicted in Figure \ref{fig:GLMmodels}.

The GLM1 and GLM2 models are utilized to simulate an accretion disk with an emission profile extending all the way towards the event horizon, being distinguished by the radial emission decay near this region. In contrast, the GLM3 model represents an accretion disk with its emission peak located near the radius of the innermost stable circular orbits ($r_{ISCO}$), which is $6M$ for a Schwarzschild black hole. The latter allows us to separate the individual luminosity contributions of the $n=1$ and $n=2$ rings from the direct emission, thereby facilitating the comparison of these features between the Schwarzschild and Cotton black holes, whereas in the former models the photon rings appear as a discrete set of boosts of luminosity superimposed on top of the disk's direct emission \cite{Johnson:2019ljv}. In our simulations, the observed luminosity is normalized with respect to its total value, to ensure both geometries are compared in identical settings. We resort to the Geodesic Rays and Visualization of IntensiTY profiles (GRAVITYp) ray-tracing code, developed by some of us.

\begin{figure*}[t!]
\subfloat{
\includegraphics[width=3.95cm,height=3.50cm]{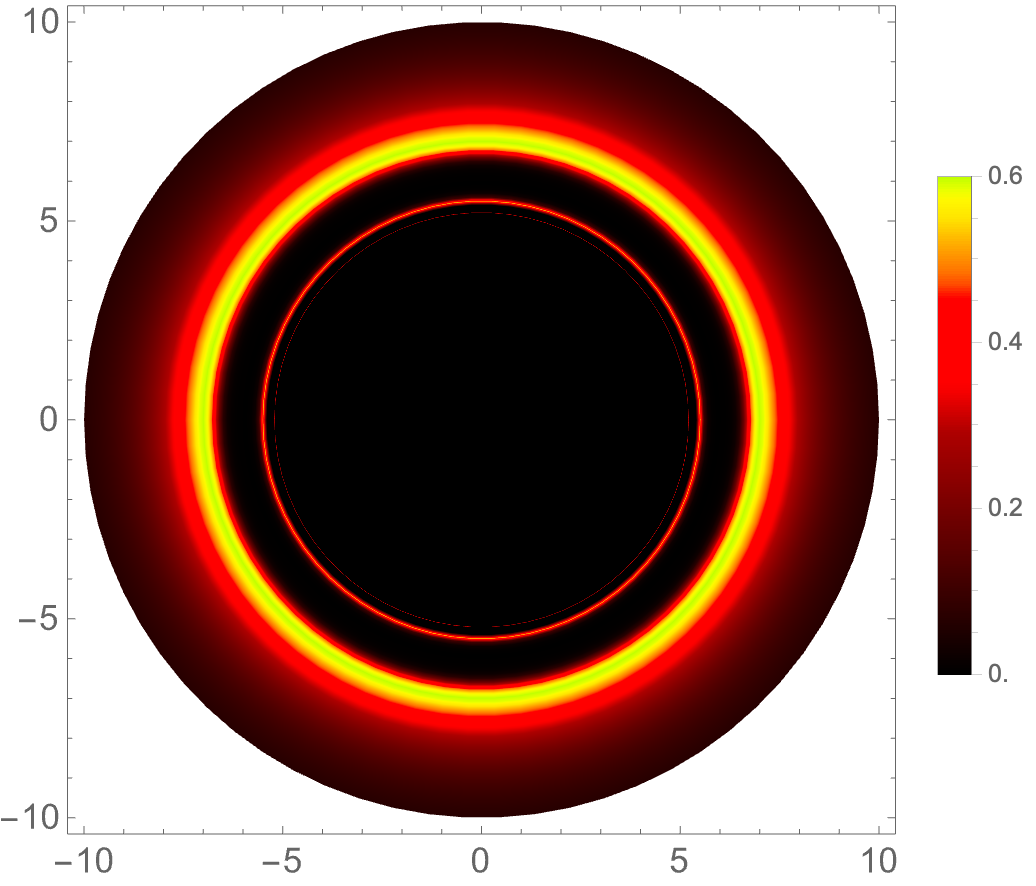}
\includegraphics[width=3.95cm,height=3.50cm]{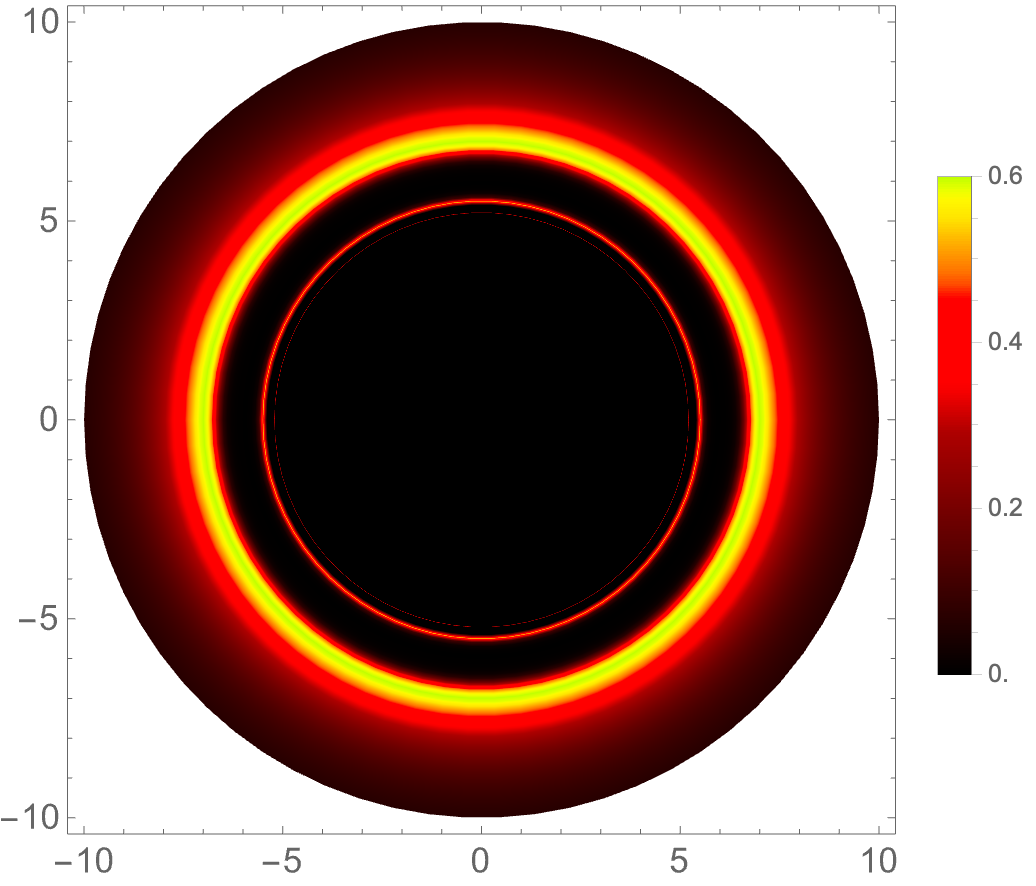}
\includegraphics[width=3.95cm,height=3.50cm]{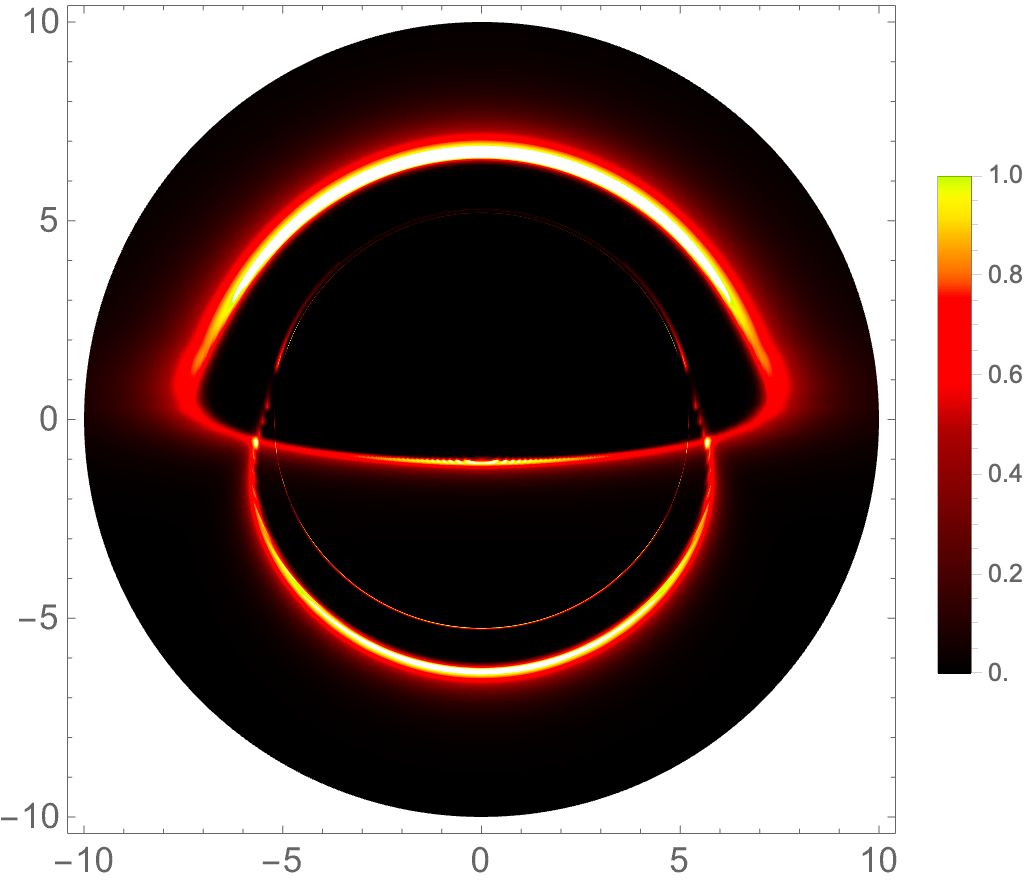}
\includegraphics[width=3.95cm,height=3.50cm]{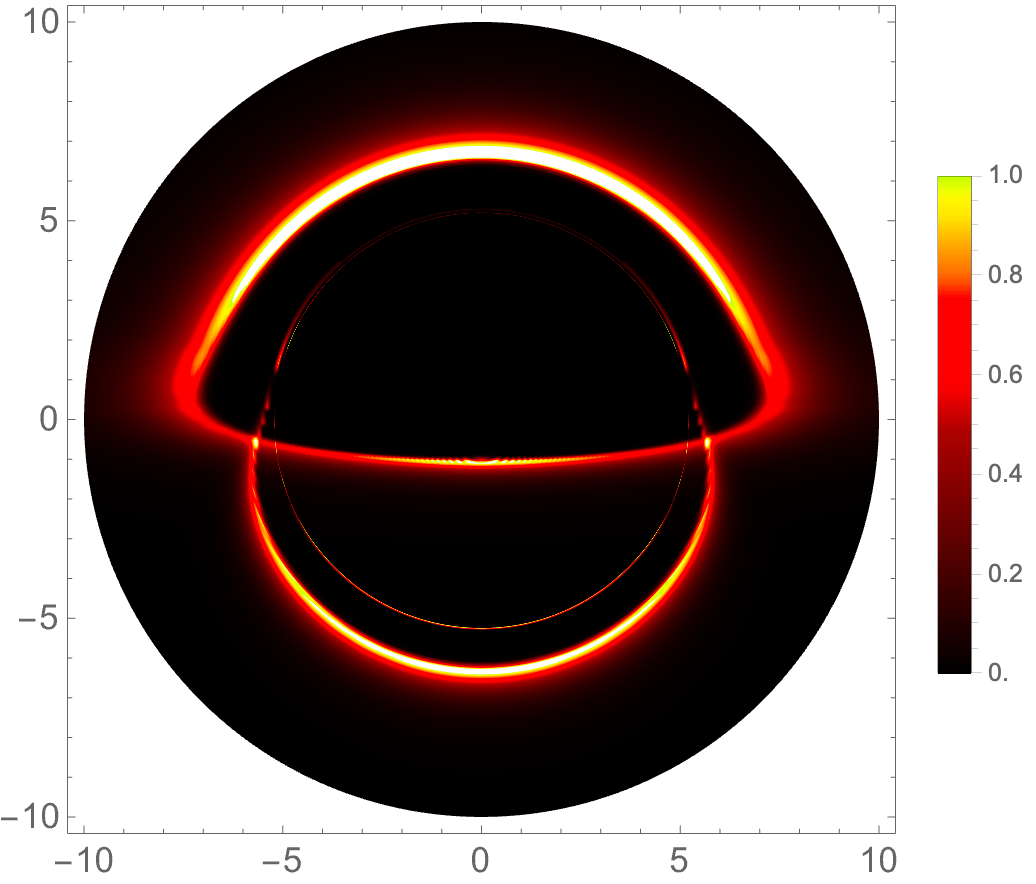}
}\quad
\subfloat{
\includegraphics[width=3.95cm,height=3.50cm]{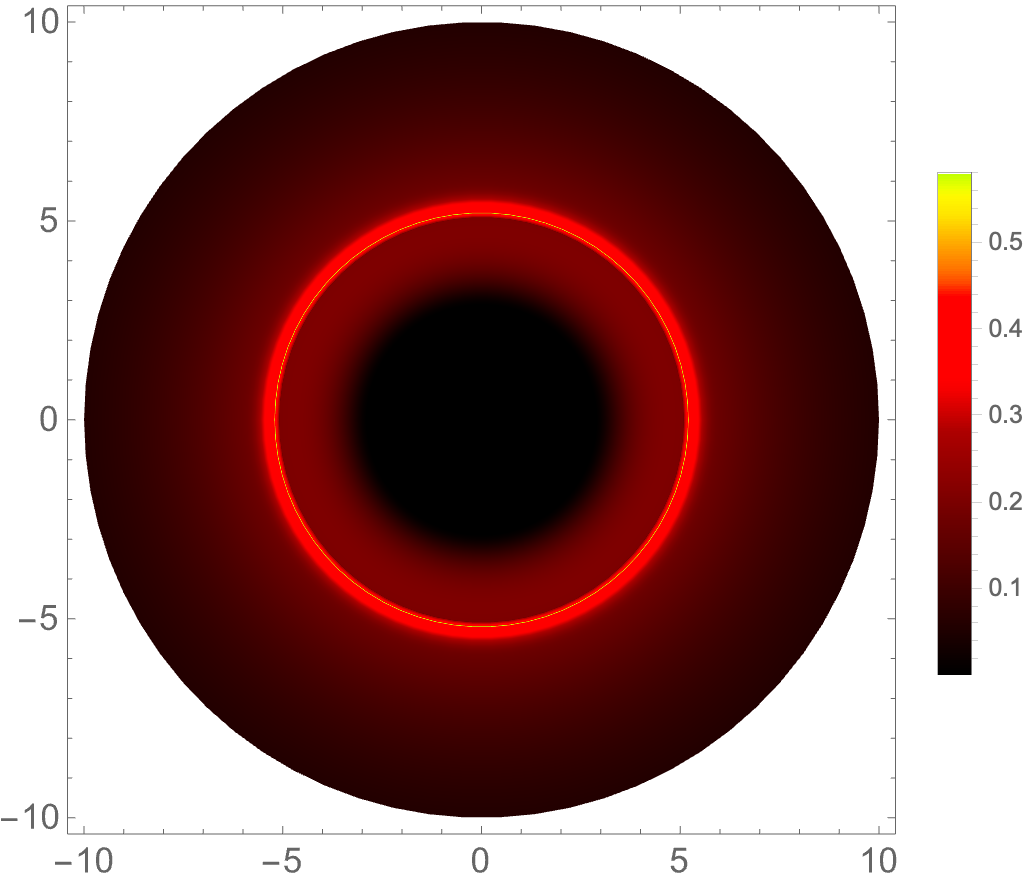}
\includegraphics[width=3.95cm,height=3.50cm]{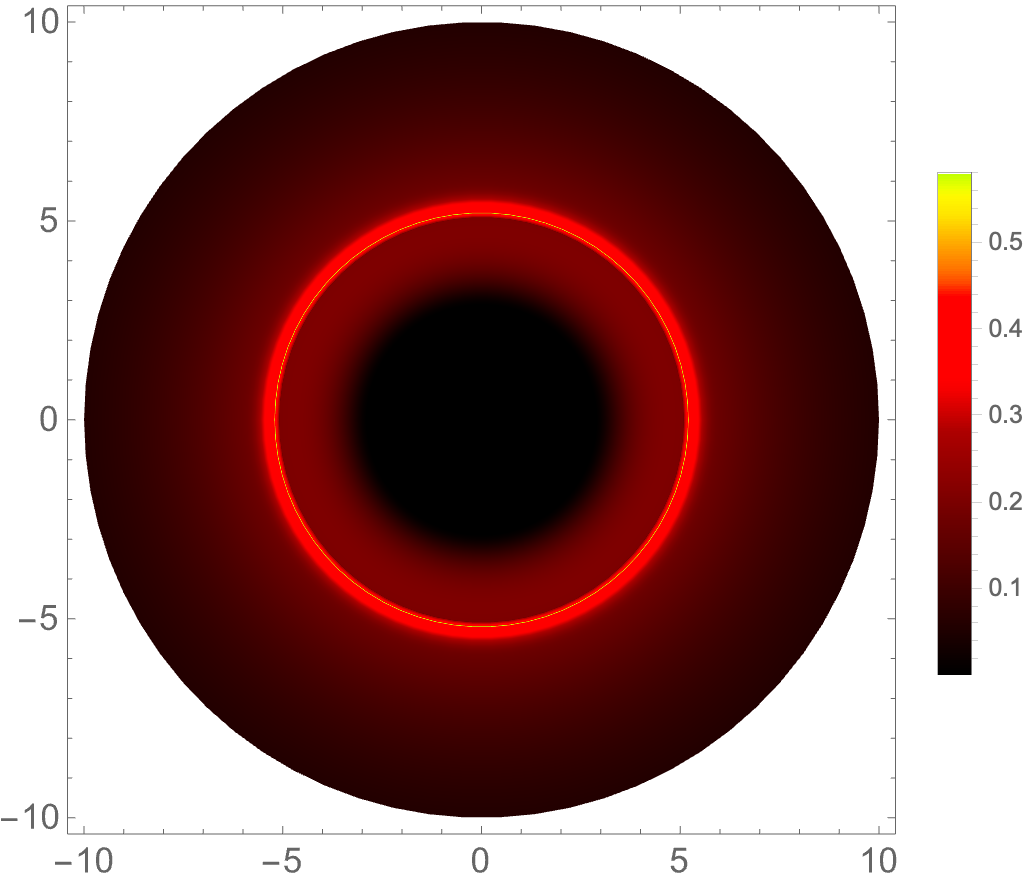}
\includegraphics[width=3.95cm,height=3.50cm]{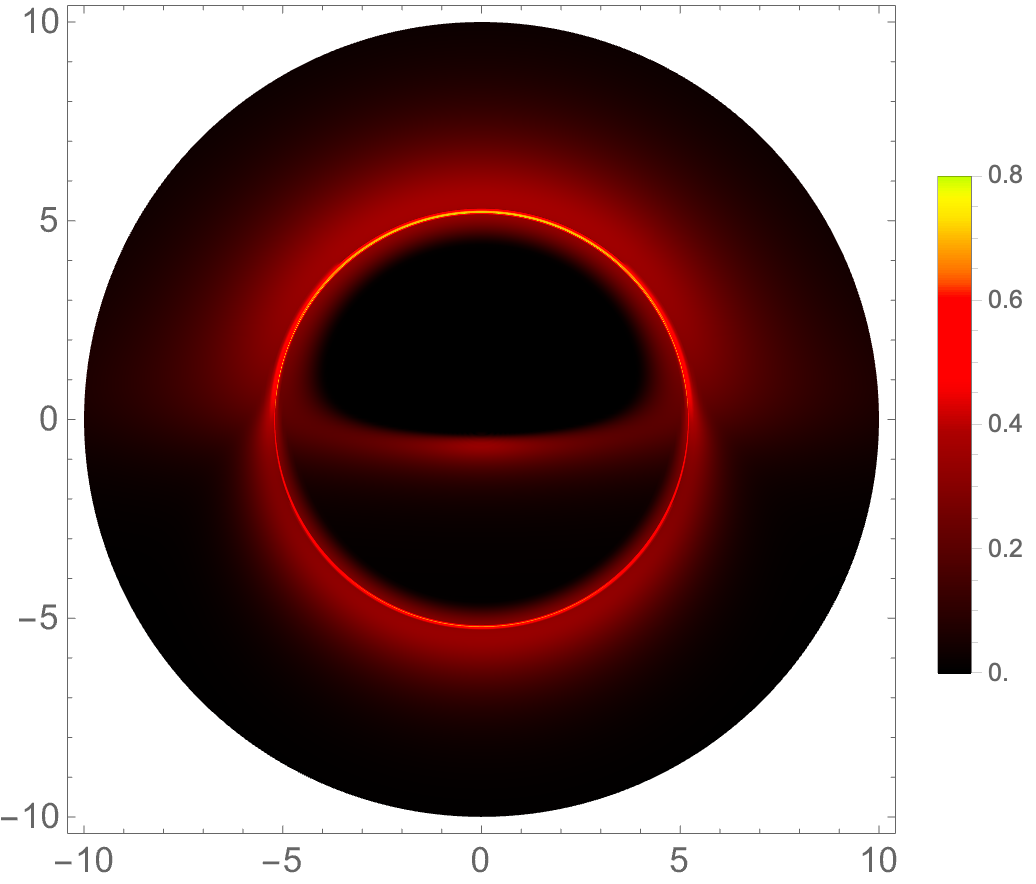}
\includegraphics[width=3.95cm,height=3.50cm]{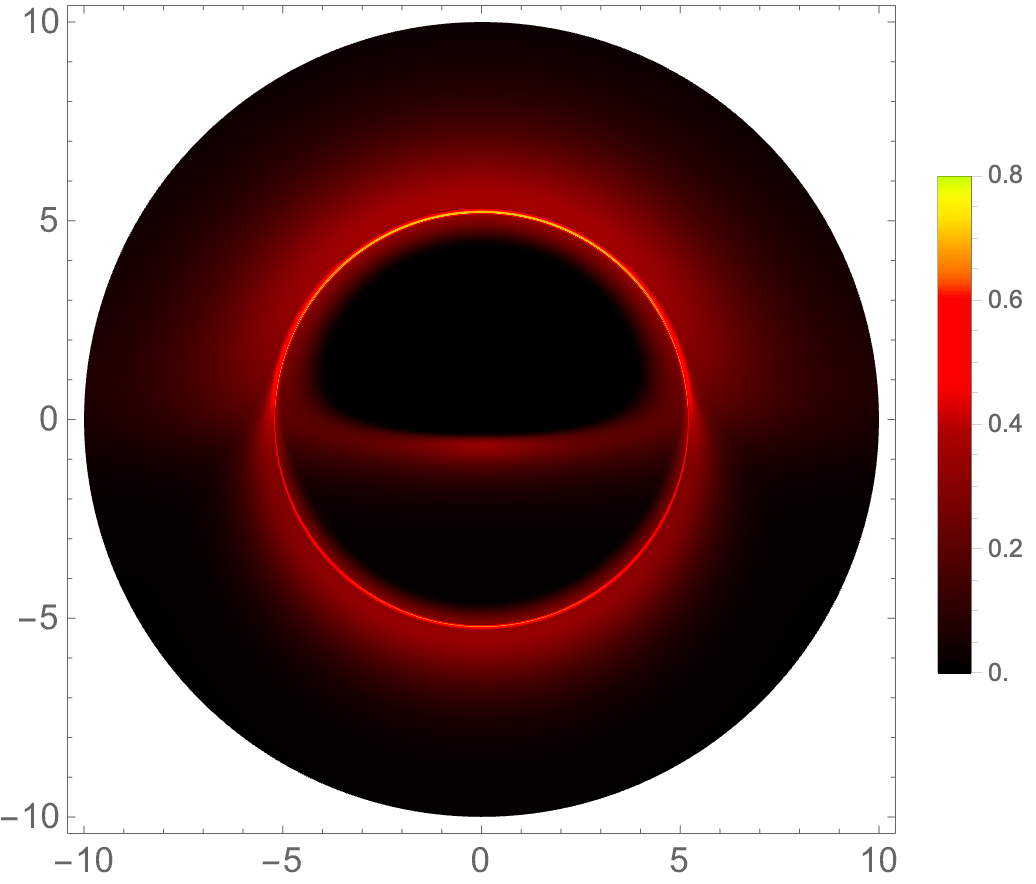}
}\quad
\subfloat{
\includegraphics[width=3.95cm,height=3.50cm]{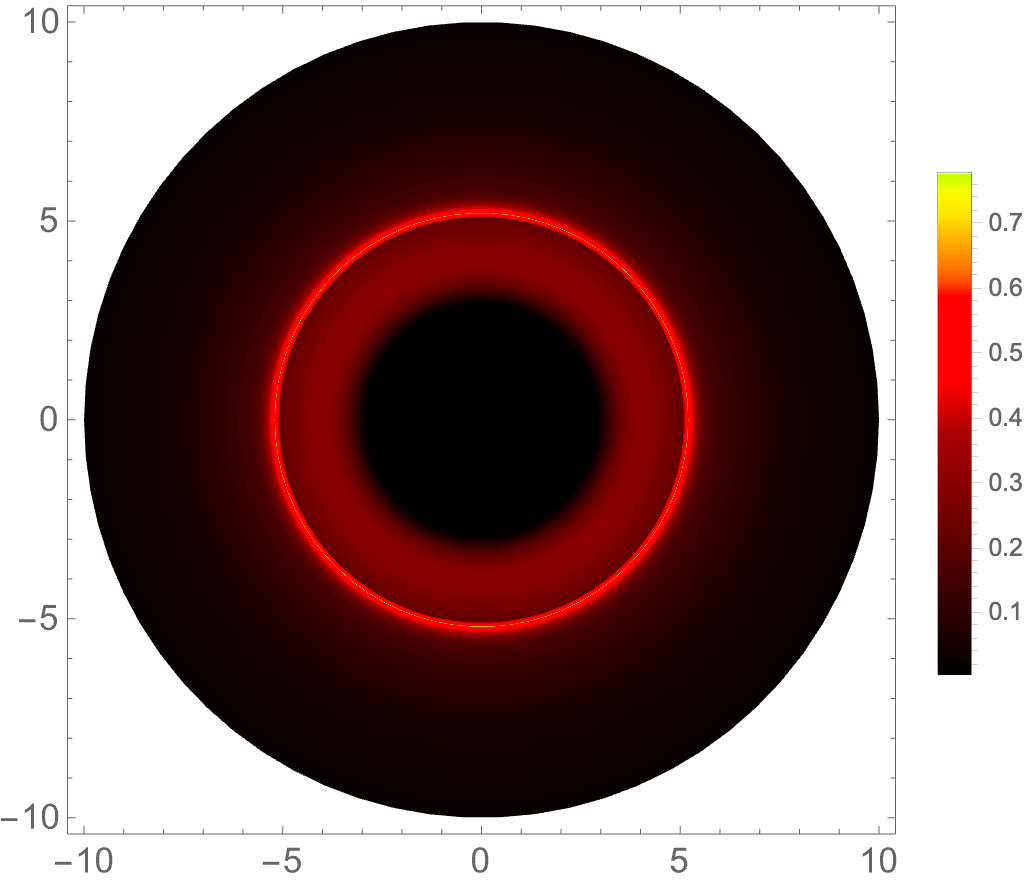}
\includegraphics[width=3.95cm,height=3.50cm]{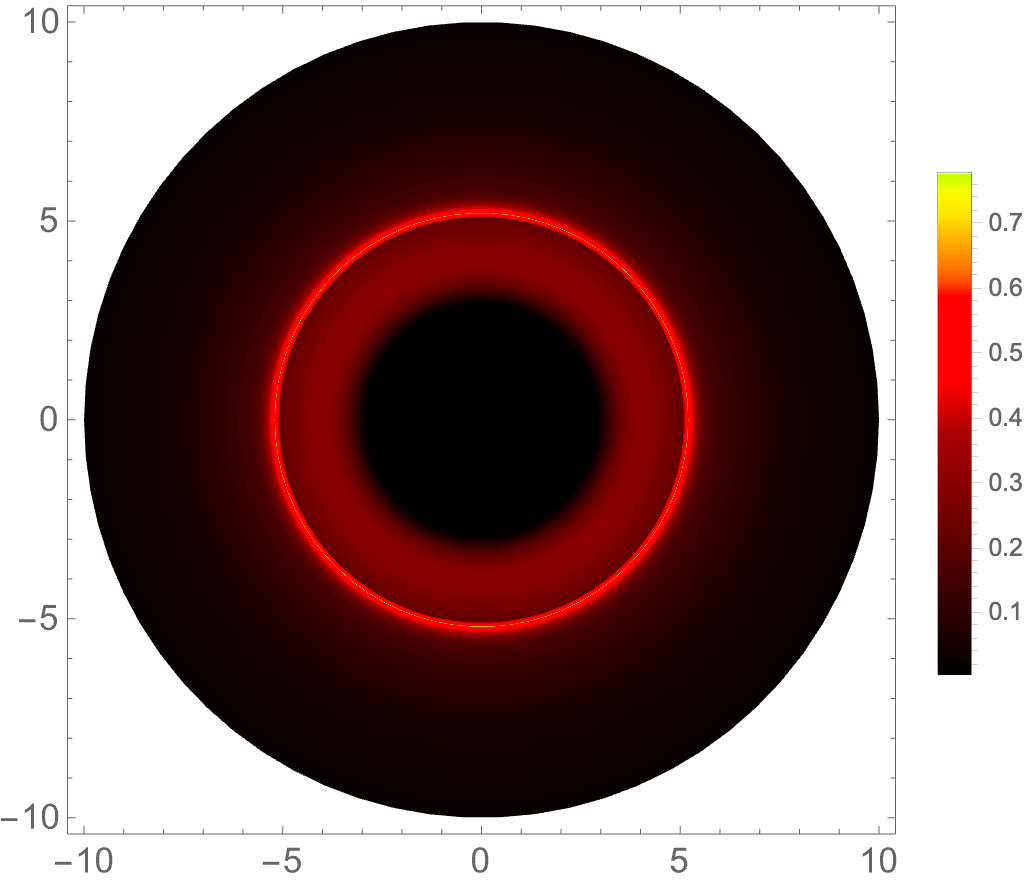}
\includegraphics[width=3.95cm,height=3.50cm]{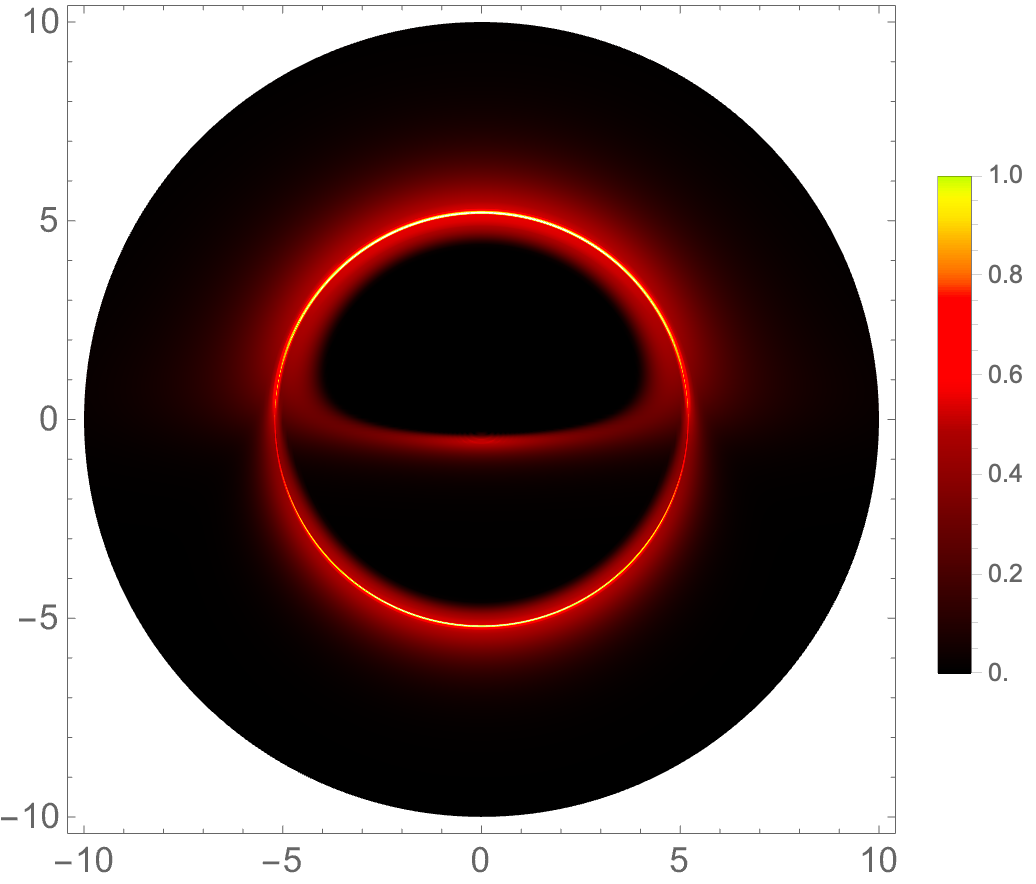}
\includegraphics[width=3.95cm,height=3.50cm]{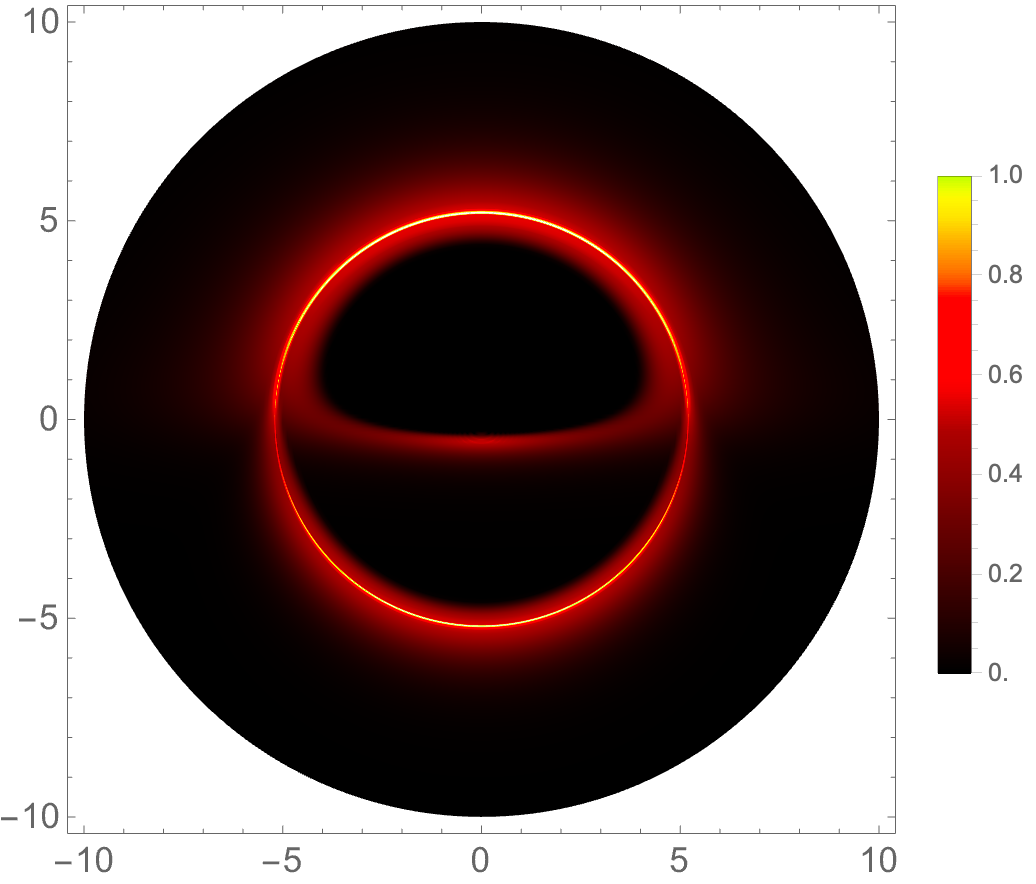}
}
\caption{Visual appearance, in the impact parameter space, of the Schwarzschild black hole ($1^{st}$ and $3^{rd}$ columns) and the CG black hole with ($\gamma M = 3.525 \times 10^{-12}$) ($2^{nd}$ and $4^{th}$ columns), for the GLM3 (top), GLM1 (middle) and GLM2 (bottom) emission profiles, for an observer located at a distance of $r_o\simeq 4.2 \times 10^{10} M$ \cite{Do:2019txf}, seen from a face-on orientation and from an inclination of $80^{\circ}$, respectively.}
\label{fig:BHimages}
\end{figure*}

By inspection of Fig. \ref{fig:BHimages}, we report no appreciable visual differences between a Schwarzschild and a Cotton black hole of the same mass, with $\gamma M \simeq 3.525 \times 10^{-12}$, regardless of the emission conditions. At the considered distance, the Cotton black hole's corresponding event horizon radius, photon sphere radius and critical impact parameter, are nearly indistinguishable from those of a Schwarzschild black hole of the same mass $M$,  differing only by a factor of the same order of magnitude of $\gamma$. Since for a non-asymptotically flat geometry the shadow radius does not coincide with the critical impact parameter, the geometries appear identical. This similarity persists onto the photon ring(s), where the gravitational lensing features play a significant role. In the literature, however,  various black holes with a similar critical impact parameter have been found to be distinguished by the properties of their corresponding photon rings \cite{daSilva:2023jxa}. These differences occur when, within the parameter space compatibility with Sgr A* shadow radius, the additional parameter in a given geometry modifies significantly the effective potential (i.e. its shape, location, concavity) seen by photons on their approach to a black hole, from the canonical Schwarzschild case. For some geometries, however, the constraints to the parameter space are too restrictive, to the point that there are virtually no changes to the effective potential, which is our case. Finally, we point out that although Eq.\eqref{eq:integratedIntensity} suggests the observed intensity depends on the observer's radial coordinate, for a non-asymptotically flat space-time, there are no visual differences between the Cotton and Schwarzschild black holes since the intensities are normalized with respect to their total value. Without normalizing the intensities, the observed intensities of the accretion disk emission in the Cotton spacetime would decrease, as in this case $A(r) \rightarrow \infty$ for an observer at infinity.

%%%%%%%%%%%%%%%%%%%%%%%%%%%%%%%%%%%%%%%%%%%%%%%%%%%%%%%%%%%%%%%%%%%%%%%%%%%%%%%%%%%%%%%%%%%%%%%%%%%%%%%%%%%%%%%%%%%%%%%%%%%%
\section{Isoradials\label{sec:IsoR}}
%%%%%%%%%%%%%%%%%%%%%%%%%%%%%%%%%%%%%%%%%%%%%%%%%%%%%%%%%%%%%%%%%%%%%%%%%%%%%%%%%%%%%%%%%%%%%%%%%%%%%%%%%%%%%%%%%%%%%%%%%%%%

In this section, we perform an analysis of isoradial curves in the Cotton and Schwarzschild space-times, to complement the black hole images presented in the previous section. These curves refer to the contours, in the observer's image plane, that lie at a radial constant distance from the black hole's center.

We solve the isoradial equatorial curves corresponding to the $n=0$ emission, following the analytical approximation provided by \cite{Claros:2024atw}, which states that the arrival point of a light ray may be mapped onto an observer's plane with Cartesian coordinates $(X',Y')$ as 

\begin{equation}
   \begin{bmatrix}
X' \\
Y' 
    \end{bmatrix}
    = \frac{R}{\sqrt{A(R)}}\sqrt{1-(1-x)^2}
    \begin{bmatrix}
\text{cos} \varphi \\
\text{sin} \varphi
    \end{bmatrix}
\end{equation}
Here, $R$ denotes a given radial distance from the origin, $x=1-\cos \alpha$,  where $\alpha$ is the emission angle, while $\varphi$ corresponds to the polar angle in the observer's plane\footnote{For a better understanding of this coordinate representation, the reader is referred to Fig. 7 of \cite{Claros:2024atw}.}. The value of $x$ may be computed via a refinement of Poutanen approximation developed in \cite{Poutanen:2019tcd} via the following formula \cite{Claros:2024atw}
\begin{widetext}\label{eq:LrayApprox}
\[
 x=A(R)y \left\{ 1 + \left(\frac{I_1}{2} + \frac{1}{6} \right)y +\frac{(315 I_1^2 -135 I_2 + 180 A^2(R) +120 I_1 +32)}{720}y^2 -\frac{e}{100}(1-A(R))y\left[ \text{ln}\left(1-\frac{y}{2}\right) +\frac{y}{2}\right] \right\},
\]
\end{widetext}
with $y=1-\cos \psi$ (with $\psi$ the azimuthal angle) being expressed by the trigonometric relation between the polar angle in the observer's plane and the observer's inclination angle $\theta_o$ via
\begin{equation}
    y=1+\frac{\text{sin}(\varphi)}{\sqrt{\text{sin}^2(\varphi)+\text{cot}^2(\theta_o)}},
\end{equation}
and where the quantities $I_1$ and $I_2$ are calculated as
\begin{equation}
   I_1=\int_{0}^{1} P(u) \  du \ , \\\\ I_2=\int_{0}^{1} P^2(u) \  du,
\end{equation}
with $P(u)=-4u^2A(u) + A(R)$, and where the variable change $r=R/u$ was considered to obtain $A(u)$.

\begin{figure}[t!]
    \centering
   \includegraphics[width=\linewidth]{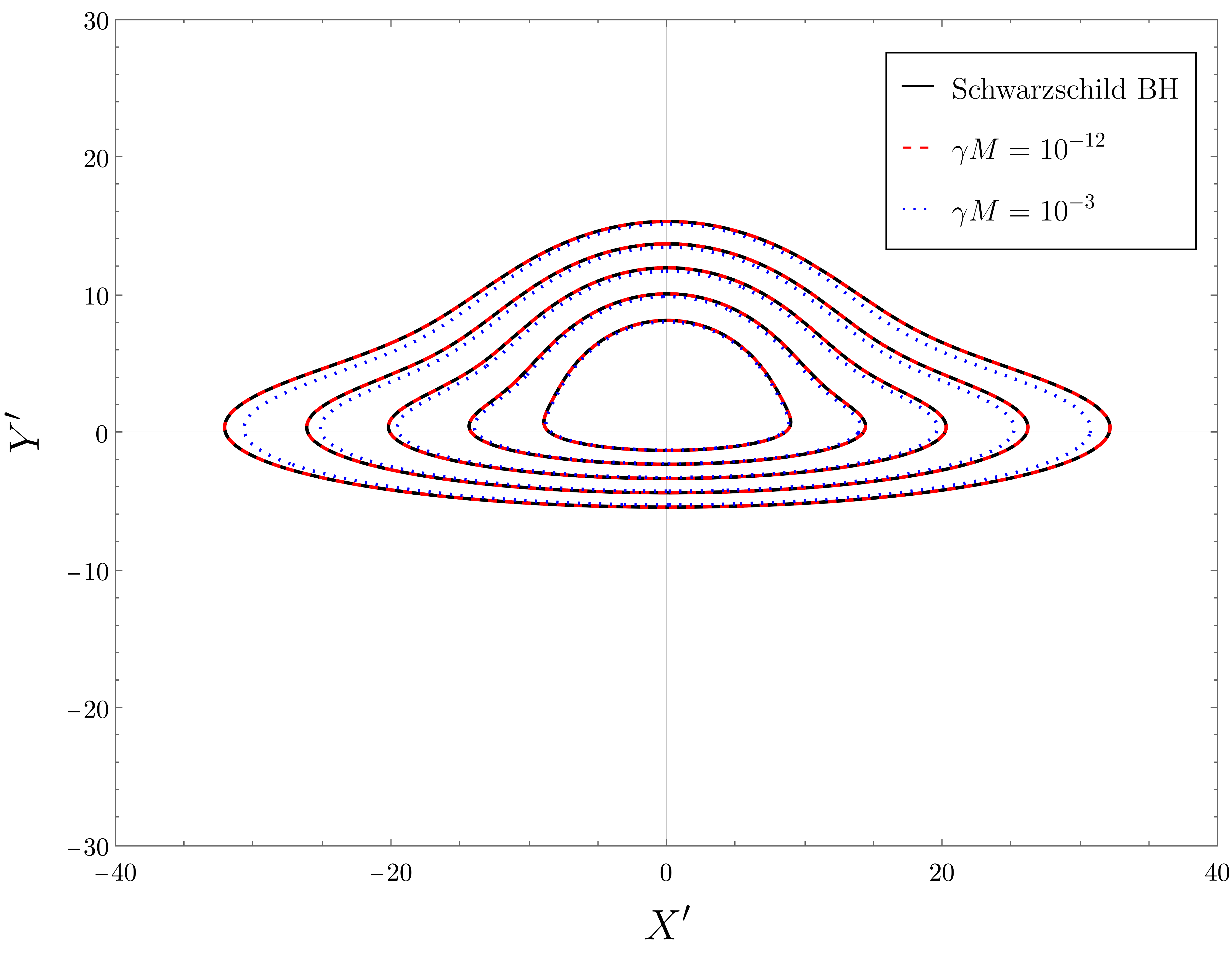}
    \caption{Isoradial curves at $R={6M, 12M, 18M, 24M, 30M}$ of a Schwarzschild black hole (black line) and a CG black hole with $\gamma M=3.525 \times 10^{-12}$ (red dashed line), as seen by a distant observer with an observation angle $\theta_o = 80^{\circ}$.}
    \label{fig:isoradials}
\end{figure}

By solving the above equations for all polar angles, for a light ray with its emission point located at a constant $r=R$, one obtains the isoradial curves of the $n=0$ emission. In Fig. \ref{fig:isoradials}, we display isoradial curves for a Schwarzschild black hole and for a Cotton black hole with $\gamma M \simeq 3.525 \times 10^{-12}$, as seen by an observer with inclination angle $\theta_o = 80^{\circ}$. 

This result emphasizes the similarity in the horizon-scale lensing phenomena between these two geometries, as seen by the complete overlap of the isoradial curve of both space-times; something which is in agreement with the analysis in the previous section. 

%%%%%%%%%%%%%%%%%%%%%%%%%%%%%%%%%%%%%%%%%%%%%%%%%%%%%%%%%%%%%%%%%%%%%%%%%%%%%%%%%%%%%%%%%%%%%%%%%%%%%%%%%%%%%%%%%%%%%%%%%%%%
\section{Theoretical Lyapunov Exponent\label{sec:Lyap}}
%%%%%%%%%%%%%%%%%%%%%%%%%%%%%%%%%%%%%%%%%%%%%%%%%%%%%%%%%%%%%%%%%%%%%%%%%%%%%%%%%%%%%%%%%%%%%%%%%%%%%%%%%%%%%%%%%%%%%%%%%%%%

\begin{figure}[t!]
    \centering
   \includegraphics[width=\linewidth]{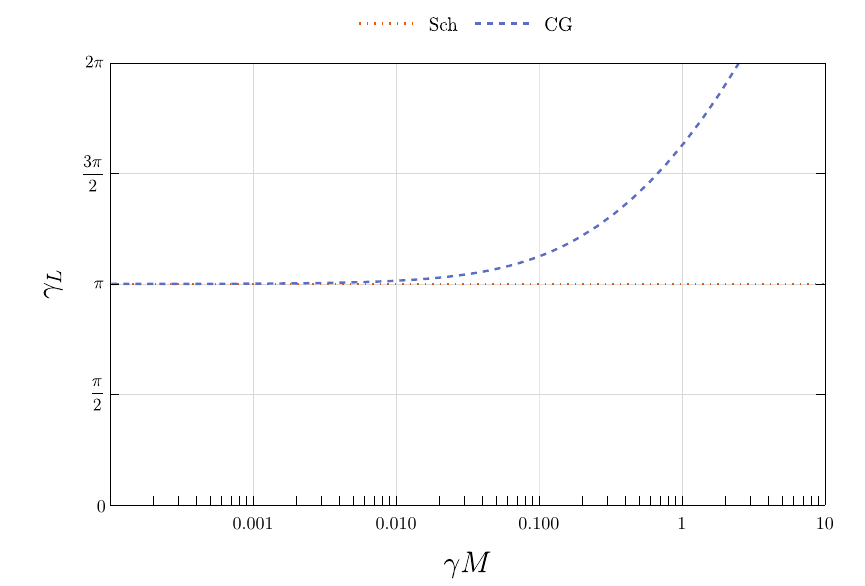}
    \caption{Lyapunov exponent as a function of $\gamma$ (blue, dashed), as described by equation \eqref{eq:lyapunov_exp}, in contrast with its Schwarzschild counterpart (red, dotted).}
    \label{fig:lyapunov}
\end{figure}

In this section we explore the Lyapunov exponent of nearly bound orbits, a quantity that measures the instability scale of such orbits. Since these instabilities take place near the photon sphere, the Lyapunov exponent tracks a local effect and is therefore independent of the observer's distance to the black hole. 
%First, we derive an expression for the theoretical Lyapunov exponent of a spherically symmetric spacetime and its relation with the number of half-orbits $n$ a photon completes around a black hole, while on its way towards a distant observer. Next, we compare the theoretical Lyapunov exponent between the Schwarzschild and Cotton spacetimes.
We begin our analysis by considering the effect of a small radial perturbation near the photon sphere (i.e. $r=\delta r + r_p \ , \delta r \ll r_p$) and apply it to Eq.\eqref{eqphi}. We perform a second-order Taylor series expansion of the effective potential $V_{eff}$ around $r_p$
\begin{equation}
    V_{eff}(r)\simeq V_{eff}(r_p)+V'_{eff}(r_p)\delta r+\frac{1}{2}V''_{eff}(r_p)(\delta r)^2,
\end{equation}
where the first term of the expansion corresponds to the critical impact parameter $b_c$ and the second term disappears, since unstable circular null geodesics correspond to critical maxima of the effective potential. The prime notation above denotes a derivative with respect to the radial coordinate. Substituting this result into Eq.\eqref{eqphi}, leading to the following expression 
\begin{equation}
    \left(\frac{d \delta r}{d\phi}\right)^2 = -\frac{1}{2}\frac{C^2(r_p)}{A(r_p)B(r_p)}V''_{eff}(r_p)(\delta r)^2.
\end{equation}
By writing the differential equation in the form
\begin{equation}
    \pi \frac{d \delta r}{d\phi} = \gamma_L \delta r,
\end{equation}
its solutions are of an exponential form, $\delta r_n=\delta r_0e^{\gamma_L n}$, where recall that $n \equiv \phi/(2\pi)$ is the number of half-turns. In this expression $\gamma_L$ is the critical exponent controlling the (exponential) drift of nearly-bound orbits with $n$, and can be traded, using the critical curve relation $C'(r_p)A(r_p)-C(r_p)A'(r_p)=0$ \cite{Claudel:2000yi}, by another exponent $\lambda$ which captures the same drift but now in coordinate time $t$ as $\gamma_L= \pi b_c \lambda$, where the such exponent reads as \cite{Cardoso:2008bp}
\begin{equation}
    \lambda=\pi\left(\frac{1}{2}\frac{AC''-A''C}{AB}\right)^{\frac{1}{2}}.
\end{equation}
For the Schwarzschild geometry $\gamma_L = \pi$, indicating that an initial perturbation grows (in radial distance) by a factor $e^{\gamma_L}$ with each half-orbit around the black hole. For the Cotton geometry, the Lyapunov exponent reads
\begin{equation}\label{eq:lyapunov_exp}
    \gamma_L=\pi(1+\gamma r_p)^{\frac{1}{2}}.
\end{equation}
In Fig. \ref{fig:lyapunov} we compare the Lyapunov exponent of the CG geometry, as described by Eq.(\ref{eq:lyapunov_exp}), against that of the Schwarzschild geometry.

This result supports the qualitative analysis done in the previous sections. The $\lambda$ exponent of the CG black hole, differs from the Schwarzschild black hole by an order of magnitude proportional to $\gamma$.  Thus, despite the exponential behaviour driven by the Lyapunov exponent, when $\gamma M \sim 10^{-12}$ the deviations between these two geometries are captured by photon rings of a similar order, requiring many half-orbits around the black hole before becoming significantly different. The gravitational lensing causing the $n=1$ and $n=2$ photon rings is far from being sensitive to the contribution of $\gamma$. Thus, one cannot hope to be able to distinguish between the Schwarzschild and CG geometries via their $n=1$ and $n=2$ ring features (i.e. relative intensities, relative widths), which are the ones that could be within reach of future interferometric detectors \cite{Ayzenberg:2023hfw}.

\section{Conclusion \label{sec:conclu}}

In this work we investigate the phenomenon of gravitational lensing and its associated observables within an extension of GR recently proposed in the literature and dubbed as Cotton gravity. We estimated the size of the new parameter $\gamma$ of Cotton gravity and which arises with a newly found spherically symmetric vacuum (Schwarzschild-like) solution, by making a numerical comparison with bound on the shadow's size of the supermassive black hole at the center of our galaxy (Sgr A*) reported by the Event Horizon Telescope Collaboration. Such a bound  is saturated, for CG, for values as small as $\gamma M = 3.525 \times 10^{-12}$ (at $2\sigma$ in such bounds), where we took into account some subtleties associated to the non-asymptotic flatness of CG geometry to identifying the mass parameter of it with the one of the Schwarzschild case. 

To study the impact of this result in geodesic effects, we first computed the exact expression for the deflection angle in the form of elliptic integrals. There are two important considerations about this result. Firstly, in order to obtain the result of the Schwarzschild solution, one must first assume $\gamma=0$ and additionally $u_R$ and $u_S=0$ (which corresponds to the consideration of infinite distances). This is because the result for Schwarzschild is usually determined using the definition for infinite distances, which is not possible in the case of Cotton gravity given its non-asymptotically flat character. Secondly, the deflection angle can be written as a function of the impact parameter $b$, though under an explicit cumbersome form. We furthermore computed the deflection angle following the approach of Bozza for the strong-field limit, searching for the roots of the deflection function $G(z)$ and verifying that the resulting expression smoothly recovers the right weak-field limit, 
where $M/b<< 1$. We have considered a linear approximation for the parameter $\gamma$ in addition to the series with respect to the mass, and checked the consistence of the analytical approximation with the numerical integration of the deflection angle. Overall, we find that the deflection angle remains unchanged for values of the CG parameter $\gamma M < 10^{-5}$.

The smallness of such a parameter of CG drove us to consider the chances of distinguishing between Schwarzschild and CG black holes using observables associated to gravitational lensing. To this end, we simulated the appearance of a CG black hole, with such a value of the $\gamma$ parameter, surrounded by a thin accretion disk, modelled by three semi-analytic radial intensity profiles. We found the constraints placed on $\gamma$ by Sgr A*'s shadow bounds to be too tight for any significant visual differences to appear. In this regard, we note that although in CG the observed intensity depends on the observer's radial distance, our images do not show any difference in this regard due to having normalized intensity profiles. 

In looking for further settings to search for the presence of any differences, we studied the isoradial curves of a CG black hole against those of a Schwarzschild black hole, utilizing the generalization developed in \cite{Claros:2024atw} of the Poutanen formula of Ref. \cite{Poutanen:2019tcd}. We found that there is a full overlap of the isoradial curves between the CG and Schwarzschild geometries, confirming the absence of any visual differences in this regard. We also considered the theoretical Lyapunov exponent for a Cotton gravity black hole, associated to the instability scale of photon rings. We derived its exact expression by perturbing the photon equation of motion about the photon sphere radius. The CG theoretical Lyapunov exponent was found to differ from Schwarzschild's by a factor proportional to CG's $\gamma$ parameter, and thus photon rings do not help in discriminating between both geometries, at least for those rings ($n=1$ and $n=2$) that could be expected to be measured in the last decades via interferometric means.

To conclude, the analysis carried out in this work points that the CG space-time recently found in the literature is very unlikely to be distinguishable from the expectations of the canonical Schwarzschild black holes with the devices we have at this moment and in the foreseeable near future. 

\acknowledgments
%%%%%%%%%%%%%%%%%%%%%%%%%%%%%%%%%%%%%%%%%%%%%%%%%%%%%%%%%%%%%%%%%%%%%%%%%%

MER thanks Conselho Nacional de Desenvolvimento Cient\'ifico e Tecnol\'ogico - CNPq, Brazil, for partial financial support. This study was financed in part by the Coordena\c{c}\~{a}o de Aperfei\c{c}oamento de Pessoal de N\'{i}vel Superior - Brasil (CAPES) - Finance Code 001.
FSNL acknowledges support from the Funda\c{c}\~{a}o para a Ci\^{e}ncia e a Tecnologia (FCT) Scientific Employment Stimulus contract with reference CEECINST/00032/2018, and funding through the research grants UIDB/04434/2020, UIDP/04434/2020 and PTDC/FIS-AST/0054/2021.
DRG is supported by the Spanish Agencia Estatal de Investigación Grant No. PID2022-138607NB-I00 and CNS2024-154444, funded by MCIN/AEI/10.13039/501100011033, FEDER, UE, and ERDF A way of making Europe.

%%%%%%%%%%%%%%%%%%%%%%%%%%%%%%%%%%%%%%%%%%%%%%%%%%%%%%%%%%%%%%%%%%%%%%%%%%


\begin{thebibliography}{999}
%%%%%%%%%%%%%%%%%%%%%%%%%%%%%%%%%%%%%%%%%%%%%%%%%%%%%%%%%%%%%%%%%%%%%%%%%%

%%%%%%%%%%%%%%%%%%%%%%%%%%%%%%%%%

%\bibitem{Cotton}
%Junpei HARADA. Emergence of the Cotton tensor for describing gravity. Physical Review D, v. 103, n. 12, p. L121502, 2021. [\href{https://journals.aps.org/prd/abstract/10.1103/PhysRevD.103.L121502}{\tt https://doi.org/10.1103/PhysRevD.103.L121502}]










%%%%%%%%%%%%%%%%%%%%%%%%%%%%%%%%%%%%%%%%%%%%%%%%%%%%%%%%%%%%%%%%%%%%%%%%%%%%%%
%%%%%%%%%%%%%%%%%%%%%%%%%%%%%%%%%%%%%%%%%%%%%%%%%%%%%%%%%%%%%%%%%%%%%%%%%%%%%% 

\bibitem{Will:2014kxa}
C.~M.~Will,
%``The Confrontation between General Relativity and Experiment,''
Living Rev. Rel. \textbf{17} (2014) 4.

%\cite{SupernovaSearchTeam:1998fmf}
\bibitem{SupernovaSearchTeam:1998fmf}
A.~G.~Riess \textit{et al.} [Supernova Search Team],
Astron. J. \textbf{116}  (1998) 1009.

%\cite{SupernovaCosmologyProject:1998vns}
\bibitem{SupernovaCosmologyProject:1998vns}
S.~Perlmutter \textit{et al.} [Supernova Cosmology Project],
Astrophys. J. \textbf{517} (1999) 565.

\bibitem{Senovilla:2022vlr}
J.~M.~M.~Senovilla,
%``The influence of Penrose\textquoteright{}s singularity theorem in general relativity,''
Gen. Rel. Grav. \textbf{54} (2022)  151.

\bibitem{Hawking:2005kf}
S.~W.~Hawking,
%``Information loss in black holes,''
Phys. Rev. D \textbf{72} (2005) 084013.

\bibitem{QG}
C. Rovelli, ``Quantum Gravity", Cambridge University Press (Cambridge Monographs on Mathematical Physics, 2004.).

%\cite{Clifton:2011jh}
\bibitem{Clifton:2011jh}
T.~Clifton, P.~G.~Ferreira, A.~Padilla and C.~Skordis,
Phys. Rept. \textbf{513}  (2012) 1.

%\cite{Capozziello:2011et}
\bibitem{Capozziello:2011et}
S.~Capozziello and M.~De Laurentis,
Phys. Rept. \textbf{509} (2011) 167.


%\cite{Harada:2021bte}
\bibitem{Harada:2021bte}
J.~Harada,
Phys. Rev. D \textbf{103}   (2021) L121502.

%\cite{Harada:2022edl}
\bibitem{Harada:2022edl}
J.~Harada,
Phys. Rev. D \textbf{106} (2022) 064044.


%\cite{Sussman:2023eep}
\bibitem{Sussman:2023eep}
R.~A.~Sussman and S.~Najera,
[arXiv:2312.02115 [gr-qc]].

%\cite{Gogberashvili:2023wed}
\bibitem{Gogberashvili:2023wed}
M.~Gogberashvili and A.~Girgvliani,
Class. Quant. Grav. \textbf{41}  no.2, 025010 (2024).

\bibitem{Mantica:2022flg}
C.~A.~Mantica and L.~G.~Molinari,
%``Codazzi tensors and their space-times and Cotton gravity,''
Gen. Rel. Grav. \textbf{55} (2023) no.4, 62

%\cite{Sussman:2023wiw}
\bibitem{Sussman:2023wiw}
R.~A.~Sussman and S.~Najera,
[arXiv:2311.06744 [gr-qc]].

\bibitem{Congdon} 
Arthur B. Congdon, and 
Charles R. Keeton, ``Principles of Gravitational Lensing: Light Deflection as a Probe of Astrophysics and Cosmology", Springer,  (2018).

%\cite{Bozza:2002zj}
\bibitem{Bozza:2002zj}
V.~Bozza,
Phys. Rev. D \textbf{66} (2002) 103001.




%\cite{Bozza:2012by}
\bibitem{Bozza:2012by}
V.~Bozza and L.~Mancini,
Astrophys. J. \textbf{753} (2012) 56.


%\cite{Pietroni:2022cur}
\bibitem{Pietroni:2022cur}
S.~Pietroni and V.~Bozza,
JCAP \textbf{12} (2022),¡ 018.


%\cite{Tsukamoto:2020iez}
\bibitem{Tsukamoto:2020iez}
N.~Tsukamoto,
Phys. Rev. D \textbf{102} (2020) 104029.


%\cite{Tsukamoto:2020bjm}
\bibitem{Tsukamoto:2020bjm}
N.~Tsukamoto,
Phys. Rev. D \textbf{103} (2021) 024033.


%\cite{Tsukamoto:2021fsz}
\bibitem{Tsukamoto:2021fsz}
N.~Tsukamoto,
Phys. Rev. D \textbf{104} (2021) 124016.



%\cite{Zhang:2022nnj}
\bibitem{Zhang:2022nnj}
J.~Zhang and Y.~Xie,
Eur. Phys. J. C \textbf{82} (2022)  471.


%\cite{Gralla:2019xty}
\bibitem{Gralla:2019xty}
S.~E.~Gralla, D.~E.~Holz and R.~M.~Wald,
Phys. Rev. D \textbf{100} (2019) 024018.

%\cite{Falcke:1999pj}
\bibitem{Falcke:1999pj}
H.~Falcke, F.~Melia and E.~Agol,
Astrophys. J. Lett. \textbf{528} (2000) L13.


%\cite{Cunha:2018acu}
\bibitem{Cunha:2018acu}
P.~V.~P.~Cunha and C.~A.~R.~Herdeiro,
Gen. Rel. Grav. \textbf{50} (2018) 42.


%\cite{Gralla:2019drh}
\bibitem{Gralla:2019drh}
S.~E.~Gralla and A.~Lupsasca,
Phys. Rev. D \textbf{101} (2020) 044031.


%\cite{Johnson:2019ljv}
\bibitem{Johnson:2019ljv}
M.~D.~Johnson,  \textit{et al.}
Sci. Adv. \textbf{6} (2020)  eaaz1310.


%\cite{EventHorizonTelescope:2019dse}
\bibitem{EventHorizonTelescope:2019dse}
K.~Akiyama \textit{et al.} [Event Horizon Telescope],
Astrophys. J. Lett. \textbf{875} (2019) L1.





%\cite{EventHorizonTelescope:2022wkp}
\bibitem{EventHorizonTelescope:2022wkp}
K.~Akiyama \textit{et al.} [Event Horizon Telescope],
Astrophys. J. Lett. \textbf{930} (2022) 12.

%\cite{Vagnozzi:2022moj}
\bibitem{Vagnozzi:2022moj}
S.~Vagnozzi, \textit{et al.}
Class. Quant. Grav. \textbf{40} (2023) 165007.


%\cite{Gralla:2020srx}
\bibitem{Gralla:2020srx}
S.~E.~Gralla, A.~Lupsasca and D.~P.~Marrone,
Phys. Rev. D \textbf{102} (2020) 124004.


\bibitem{Mannheim:1988dj}
P.~D.~Mannheim and D.~Kazanas,
%``Exact Vacuum Solution to Conformal Weyl Gravity and Galactic Rotation Curves,''
Astrophys. J. \textbf{342} (1989) 635.

\bibitem{Ghosh:2015cva}
S.~G.~Ghosh, L.~Tannukij and P.~Wongjun,
%``A class of black holes in dRGT massive gravity and their thermodynamical properties,''
Eur. Phys. J. C \textbf{76} (2016) 119.

\bibitem{Soroushfar:2015wqa}
S.~Soroushfar, R.~Saffari, J.~Kunz and C.~L\"ammerzahl,
%``Analytical solutions of the geodesic equation in the spacetime of a black hole in f(R) gravity,''
Phys. Rev. D \textbf{92} (2015) 044010.

\bibitem{Kiselev:2002dx}
V.~V.~Kiselev,
%``Quintessence and black holes,''
Class. Quant. Grav. \textbf{20} (2003) 1198.


\bibitem{Grumiller:2010bz}
D.~Grumiller,
%``Model for gravity at large distances,''
Phys. Rev. Lett. \textbf{105} (2010), 211303
[erratum: Phys. Rev. Lett. \textbf{106} (2011).


%\cite{Perlick:2021aok}
\bibitem{Perlick:2021aok}
V.~Perlick and O.~Y.~Tsupko,
Phys. Rept. \textbf{947} (2022) 1.



\bibitem{Do:2019txf}
T.~Do, \textit{et al.}
Science \textbf{365} (2019) 664.


%\cite{EventHorizonTelescope:2022xqj}
\bibitem{EventHorizonTelescope:2022xqj}
K.~Akiyama \textit{et al.} [Event Horizon Telescope],
Astrophys. J. Lett. \textbf{930} (2022) L17.

\bibitem{EventHorizonTelescope:2021dqv}
P.~Kocherlakota \textit{et al.} [Event Horizon Telescope],
%``Constraints on black-hole charges with the 2017 EHT observations of M87*,''
Phys. Rev. D \textbf{103} (2021) 104047.

\bibitem{Ghosh:2022kit}
S.~G.~Ghosh and M.~Afrin,
%``An Upper Limit on the Charge of the Black Hole Sgr A* from EHT Observations,''
Astrophys. J. \textbf{944} (2023) 174.

\bibitem{Afrin:2021wlj}
M.~Afrin and S.~G.~Ghosh,
%``Testing Horndeski Gravity from EHT Observational Results for Rotating Black Holes,''
Astrophys. J. \textbf{932} (2022) 51.

\bibitem{Afrin:2022ztr}
M.~Afrin, S.~Vagnozzi and S.~G.~Ghosh,
%``Tests of Loop Quantum Gravity from the Event Horizon Telescope Results of Sgr A*,''
Astrophys. J. \textbf{944} (2023) 149.

\bibitem{Afrin:2021imp}
M.~Afrin, R.~Kumar and S.~G.~Ghosh,
%``Parameter estimation of hairy Kerr black holes from its shadow and constraints from M87*,''
Mon. Not. Roy. Astron. Soc. \textbf{504} (2021)  5927.


%\cite{Arnowitt:1962hi}
\bibitem{Arnowitt:1962hi}
R.~L.~Arnowitt, S.~Deser and C.~W.~Misner,
Gen. Rel. Grav. \textbf{40} (2008) 1997.

%\cite{Xia:2024tps}
\bibitem{Xia:2024tps}
P.~Xia, D.~Zhang, X.~Ren, B.~Wang and Y.~C.~Ong,
[arXiv:2405.07209 [astro-ph.CO]].

%\cite{Ishihara:2016vdc}
\bibitem{Ishihara:2016vdc}
A.~Ishihara, Y.~Suzuki, T.~Ono, T.~Kitamura and H.~Asada,
Phys. Rev. D \textbf{94} (2016) 084015.


\bibitem{Gold:2020iql}
R.~Gold, \textit{et al.}
%``Verification of Radiative Transfer Schemes for the EHT,''
Astrophys. J. \textbf{897} (2020) 148.

%\cite{daSilva:2023jxa}
\bibitem{daSilva:2023jxa}
L.~F.~D.~da Silva, F.~S.~N.~Lobo, G.~J.~Olmo and D.~Rubiera-Garcia,
Phys. Rev. D \textbf{108} (2023) 084055.


%\cite{Claros:2024atw}
\bibitem{Claros:2024atw}
J.~Claros and E.~Gallo,
%``Accurate analytical modeling of light rays in spherically symmetric spacetimes: Applications in the study of black hole accretion disks and polarimetry,''
Phys. Rev. D \textbf{109} (2024) 124055.

\bibitem{Poutanen:2019tcd}
J.~Poutanen,
%``Accurate analytic formula for light bending in Schwarzschild metric,''
Astron. Astrophys. \textbf{640} (2020) A24.


%\cite{Claudel:2000yi}
\bibitem{Claudel:2000yi}
C.~M.~Claudel, K.~S.~Virbhadra and G.~F.~R.~Ellis,
J. Math. Phys. \textbf{42} (2001) 818.

\bibitem{Cardoso:2008bp}
V.~Cardoso, A.~S.~Miranda, E.~Berti, H.~Witek and V.~T.~Zanchin,
%``Geodesic stability, Lyapunov exponents and quasinormal modes,''
Phys. Rev. D \textbf{79} (2009) no.6, 064016

\bibitem{Ayzenberg:2023hfw}
D.~Ayzenberg, \textit{et al.}
%``Fundamental Physics Opportunities with the Next-Generation Event Horizon Telescope,''
[arXiv:2312.02130 [astro-ph.HE]].






\end{thebibliography}
\end{document}